\documentclass[12pt,epsfig,epsf]{article}
\usepackage{amsmath}
\usepackage{cite}
\usepackage{slashed}
\usepackage{epsf,color,colordvi}
\usepackage{graphicx}
\usepackage{amsmath}
\usepackage{amssymb}
\usepackage{enumerate}
\usepackage{epsfig}
\usepackage{cite}
\usepackage{fontenc}
\usepackage{float}
\usepackage[lofdepth,lotdepth,caption=false]{subfig}
\usepackage{xcolor}
\usepackage{booktabs}
\usepackage{tabularx}
\usepackage{multirow}
\usepackage{longtable}
\usepackage{adjustbox}
\usepackage{lscape}
\usepackage{dcolumn}
\usepackage{rotating}
\usepackage[english]{babel}
\usepackage[autostyle]{csquotes}
\usepackage{hyperref}
\usepackage[normalem]{ulem}
\usepackage{color}
\newcommand {\ignore}[1]{}
\definecolor{darkgreen}{cmyk}{1,0,1,0.4}
\definecolor{brown}{cmyk}{0,0.8,1,0.2}
\definecolor{darkred}{cmyk}{0,1,1,0.2}

\newcommand{\beq}{\begin{equation}}
\newcommand{\eeq}{\end{equation}}
\newcommand{\bea}{\begin{eqnarray}}
\newcommand{\eea}{\end{eqnarray}}

\newcommand{\ldm}{\Delta m_{31}^2}
\newcommand{\sdm}{\Delta m_{21}^2}

\hypersetup{
  colorlinks,
  citecolor=red,
  linkcolor=blue,
  urlcolor=blue}

\begin{document}
\begin{titlepage}

\vspace*{-3.cm}
\begin{flushright}

\end{flushright}

\renewcommand{\thefootnote}{\fnsymbol{footnote}}
\setcounter{footnote}{-1}

{\begin{center}
{\large\bf 
Majorana CP Violation Insights from Decaying Neutrinos   
\\[0.2cm]
}
\end{center}}

\renewcommand{\thefootnote}{\alph{footnote}}

\vspace*{.8cm}
\vspace*{.3cm}
{
\begin{center} 

   {\sf Sabila Parveen$^{\ast}$\,\footnote[1]{\makebox[1.cm]{Email:} sabila41\_sps@jnu.ac.in}
                }
        {\sf Soumya Bonthu$^{\dagger}$\,\footnote[2]{\makebox[1.cm]{Email:} bsoumya@okstate.edu}
        }
            {\sf Newton Nath$^{\ddagger,\,\S}$\,\footnote[3]{\makebox[1.cm]{Email:} nnath.phy@iitbhu.ac.in}        
                }     
                           {\sf Ujjal Kumar Dey$^{\P}$\,\footnote[4]{\makebox[1.cm]{Email:} ujjal@iiserbpr.ac.in}        
                }            
and                 
            {\sf Poonam Mehta$^{\ast}$\,\footnote[5]{\makebox[1.cm]{Email:} pm@jnu.ac.in}
}
\end{center}
}
\vspace*{0cm}
{\it 
\begin{center}
$^{\ast}$\,School of Physical Sciences, Jawaharlal Nehru University, 
      New Delhi 110067, India  \\
$^\dagger$\,Department of Physics, Oklahoma State University, Stillwater, OK, 74078, USA  \\
$^\ddagger$\,AHEP Group, Institut de F\'{i}sica Corpuscular --
  CSIC/Universitat de Val\`{e}ncia, Parc Cient\'ific de Paterna.
 C/ Catedr\'atico Jos\'e Beltr\'an, 2 E-46980 Paterna (Valencia), Spain \\ $^\S$\,Department of Physics, Indian Institute of Technology (BHU), Varanasi 221005, India \\
 $^\P$\,Department of Physical Sciences, Indian Institute of Science Education and Research Berhampur, Transit Campus, Government ITI, Berhampur 760010, Odisha, India
\end{center}
}

{\Large 
\bf
 \begin{center} Abstract  
  
\end{center} 
 }

It is well-known that within the standard three flavor neutrino oscillation formalism, the Majorana phases appearing in the neutrino mixing matrix cannot have any effect on neutrino oscillation probabilities thereby evading testability at neutrino oscillation experiments.
We consider an effective non-Hermitian Hamiltonian describing three flavor neutrino oscillations with the possibility of neutrino decay and demonstrate that the two Majorana phases can entangle with the off-diagonal decay terms and appear at the level of  oscillation probabilities. Using the Cayley-Hamilton theorem, we derive approximate analytical expressions for three flavor neutrino oscillation probabilities in the presence of neutrino decay, taking into account matter effects. In the context of a long baseline neutrino experiment, we then analyse the impact of Majorana phases  on the oscillation probabilities for different channels as well as on observables related to CP violation effects in neutrino oscillations. Finally, we discuss the effect of Majorana phases on the parameter degeneracies in the neutrino oscillation framework.

\vspace*{.5cm}

\end{titlepage}

\section{Introduction}
\label{sec:introduction}
The discovery of neutrino oscillations has confirmed that neutrinos possess tiny, non-zero masses and undergo mixing~\cite{nobel2015} (see  Table~\ref{tab:parameters} and Ref.~\cite{Capozzi:2017ipn, deSalas:2020pgw,  Esteban:2024eli, ParticleDataGroup:2024cfk} for global analyses of neutrino oscillation data). 
This observational evidence provides the highest impetus to searches for the physics beyond the Standard Model (SM) as neutrinos are known to be  massless in the SM. 
Despite the remarkable success of neutrino oscillation experiments, several fundamental questions (such as nature of the neutrino, absolute mass of neutrino, smallness of the neutrino mass in comparison to other fermions) remain unresolved.
 Dedicated experiments are currently underway to tackle these issues and to address the challenging question of determination of the nature of neutrinos~\cite{Case:1957zza, Li:1981um, Kayser:1981nw, Kayser:1982br, Kim:2021dyj, Barger:2002vy, Pascoli:2002qm, deGouvea:2002gf, Pascoli:2005zb, Simkovic:2012hq}. In particular, testing the Majorana nature of neutrinos is expected to come from lepton number-violating processes, such as neutrinoless double beta decay experiments~\cite{Adams:2022jwx} (also see~\cite{Dery:2024lem} for subtleties).
Recently, it has been suggested that neutrinos at non-relativistic energies such as the cosmic neutrino background  could allow us to determine  the nature of neutrinos~\cite{Akhmedov:2024fsh}.
The effective  electron anti-neutrino mass is constrained to be less than $0.45$ eV at 90\% C.L. from KATRIN~\cite{KATRIN:2024cdt}. There are bounds on the sum of neutrino masses from cosmological observations as well. Within the $\Lambda$CDM model, the latest DESI fit to BAO+CMB measurements constrain $\sum m_{\nu} < 0.072$ eV~\cite{DESI:2024mwx}.

Neutrinos hold a distinct position among it SM peers as they are the electrically neutral elementary particles.
In principle, this allows them to be their own anti-particles. For massive neutral fermions, we can write down the mass term in two  ways: Dirac-type (similar to other fermions) or Majorana-type (unique to neutrinos). In addition, the mixing matrix carries the imprint of nature of neutrino via additional phases in case of Majorana neutrinos. In general, for Dirac neutrinos, we have ${n(n-1)}/2$ angles and ${(n-1)(n-2)}/2$ phases to parametrize the mixing matrix. For Majorana neutrinos, we have ${n(n-1)}/2$ angles and ${n(n-1)}/2$ phases.   For the Dirac case, the Lagrangian conserves global $U(1)$ symmetry resulting in mixing of $n$ Dirac fields to give $(n-1)(n-2)/2$ physical phases. Usually, such $U(1)$ symmetry leads to the conservation of  quantum numbers such as, electric charge, lepton number etc.
For Majorana case,  additional terms are allowed in the Lagrangian that break the aforementioned global $U(1)$ symmetry. Consequently, the quantum numbers such as lepton number could be violated by two units. Thus, for $n$ Majorana fields, additional $(n-1)$ number of phases appear, referred to as the Majorana phases~\cite{Bilenky:1980cx, Schechter:1980gr, Doi:1980yb, Giunti:2010ec}.
 This implies that even for two flavor case ($n=2$), we have one Majorana phase in the mixing matrix which could be responsible for CP violation effects. For three flavor case ($n=3$), we have two Majorana phases over and above the Dirac-type CP phase.
It turns out that neutrino oscillation probability is, however, insensitive to the Majorana phases~\cite{Bilenky:1980cx, Schechter:1980gr, Doi:1980yb, Giunti:2010ec} for propagation in vacuum or in matter (with SM interactions) and therefore it is generally believed that it is impossible to probe the nature of neutrinos via neutrino oscillations experiments.

However, for certain beyond the SM scenarios, it becomes viable to see the imprints of  Majorana phases at the level of oscillation probabilities. The first such proposition came in~\cite{Benatti:2001fa} wherein  quantum decoherence effects were taken into account and under certain choice of parameters (non-zero off-diagonal elements of the decoherence matrix), it was shown that  the oscillation probability for two neutrino flavor had explicit dependence on the Majorana phase. Thus, quantum decoherence effects allowed for distinguishing between the Dirac and Majorana neutrinos for two~\cite{Capolupo:2018hrp, Capolupo:2020hqm} or three flavors~\cite{Carrasco-Martinez:2020mlg, Buoninfante:2020iyr} of neutrinos. 
It turns out that in the presence of decay, the effective Hamiltonian becomes non-Hermitian and the two-flavor neutrino oscillation probability may depend on the Majorana phase~\cite{Dixit:2022izn} for certain choice of parameters, thereby allowing for yet another possibility to distinguish between Dirac and Majorana neutrinos.
It is shown in~\cite{Soni:2023njf} (see also~\cite{Shafaq:2021lju}) that the temporal correlations in the form of Leggett-Garg inequalities carried the imprint of the nature of neutrinos for the case of two-flavor neutrino oscillations along with decay.
Strong bounds from neutrinoless double beta decay~\cite{Ge:2016tfx} can also reveal the nature of neutrinos.

Neutrino decay is typically characterized by the term, $\exp [ -(m_i/\tau_i) (L/E) ]$
where $m_i$ and $\tau_i$ are the mass and lifetime of the neutrino mass eigenstate $\nu_i$. This term gives an idea of the fraction of neutrinos of energy $E$ that decay after traversing a distance $L$. Several neutrino observations have led to constraints on the invisible decay parameters. 
The observation of supernova neutrinos from SN1987A leads to the constraint $\tau/m > 10^{5}$ s/eV for at least one of the mass eigenstates of neutrinos~\cite{Frieman:1987as}.  Analysis of oscillation plus decay solution in the context of MINOS and T2K~\cite{Gomes:2014yua} and in the context of NO$\nu$A and T2K data~\cite{Choubey:2018cfz} constrain $\tau_3/m_3 > 2.8 \times 10^{-12}$ s/eV ($90\%$ C.L.) and $\tau_3/m_3 > 1.5 \times 10^{-12}$ s/eV $(3\sigma)$, respectively.
Recent combined analysis of the data sets from the MINOS/MINOS+, NO$\nu$A and T2K experiments constrain  $\tau^{}_{3}/m^{}_{3}\geq 2.4 \times 10^{-11}\,\rm{s/eV}$ at $90\%$ C.L.~\cite{Ternes:2024qui}. 
From SNO, the limit on $\nu_2$ lifetime is $\tau^{}_{2}/m^{}_{2} > 8.7 \times 10^{-5}\,\rm{s/eV}$~\cite{Bandyopadhyay:2002qg}. 
From solar neutrino data, a lower bound on the lifetime has been obtained for the $\nu^{}_{2}$ state,  $\tau^{}_{2}/m^{}_{2}>7.2 \times 10^{-4}\,\rm{s/eV}$ at $99\%$ C.L.~\cite{Picoreti:2015ika}. 
A combination of atmospheric data from SK, and long baseline data from K2K and MINOS leads to the constraint, $\tau_3/m_3 \ge 9.3 \times 10^{-11}$ s/eV at 99\% C.L.
~\cite{Gonzalez-Garcia:2008mgl}.

The current and upcoming neutrino oscillation experiments such as, T2K~\cite{T2K:2017hed}, NO$\nu$A \cite{NOvA:2016kwd}, JUNO~\cite{JUNO:2015zny}, T2HK~\cite{Hyper-KamiokandeProto-:2015xww,Hyper-Kamiokande:2018ofw}, DUNE~\cite{Acciarri:2015uup, DUNE:2020ypp}  as well as the proposed experiments such as T2HKK~\cite{Hyper-Kamiokande:2016srs} and ESS$\nu$SB~\cite{ESSnuSB:2013dql} have the potential to explore physics beyond the SM such as neutrino decay  which could give rise to sub-dominant effects.
Invisible decay with oscillation in the context of future atmospheric neutrino experiments
have been studied for INO~\cite{Choubey:2017eyg} and for KM3Net-ORCA~\cite{deSalas:2018kri}. Atmospheric neutrinos in conjunction with an iron calorimeter can constrain $\tau^{}_{3}/m^{}_{3}>1.51 \times 10^{-10}\,\rm{s/eV}$ at $90\%$ C.L. for an exposure of 500 kt-yr~\cite{Choubey:2017eyg}. Reactor neutrino oscillation experiments with medium-baseline such as JUNO (or RENO) can place constraints on the $\nu^{}_{3}$ state, $\tau^{}_{3}/m^{}_{3}>7.5\,(5.5) \times 10^{-11}\,\rm{s/eV}$ at $95\%\,(99\%)$ C.L. for an exposure of 100 kt-yr~\cite{Abrahao:2015rba}. 
 Invisible neutrino decay at neutrino telescopes has been explored in~\cite{Beacom:2002cb, Pakvasa:2012db, Denton:2018aml}.
Strong constraints on the lifetime of invisible neutrino decay can be derived from cosmology~\cite{Hannestad:2005ex,Basboll:2008fx,Escudero:2019gfk,Barenboim:2020vrr,Chen:2022idm}.

Neutrino decay could be either visible or invisible. 
In what follows, we shall consider invisible decays, i.e., the final states remain undetected. 
It should be pointed out that neutrino decay was invoked as an explanation for various anomalies, however, presently it is disfavoured as a leading mechanism of flavor conversion and can only be considered as a sub-leading effect~\cite{Bahcall:1972my, Bahcall:1986gq, LoSecco:1998cd}.
%
Various models have been studied in the context of neutrino decay scenarios. One such model involves the Majoron, a light or massless particle, possibly related to neutrino mass generation~\cite{Gelmini:1980re, Chikashige:1980ui,Schechter:1981cv}. Other models include mirror fermion~\cite{Maalampi:1988vs}, supersymmetry ~\cite{Gabbiani:1990uc,Enqvist:1992ef,Aboubrahim:2013gfa}, left-right symmetric theories~\cite{Kim:2011ye}, and topological approaches linked to gravitational anomalies~\cite{Dvali:2016uhn}. These models offer different ways to understand and test neutrino decay in experiments.

In the past, the analytic treatment of neutrino oscillation plus decay for two or three flavor scenario relied on the assumption that the neutrino mass eigenstates and the decay eigenstates coincide~\cite{Lindner:2001fx, Abrahao:2015rba, Ghoshal:2020hyo}. While this could be true for specific scenarios in vacuum, in general the propagation in matter is expected to induce a mismatch between these eigenstates. Recently, a proper analytic treatment of neutrino oscillation plus decay for the general case and including the effects due to Earth matter has been put forth for two flavor case~\cite{Chattopadhyay:2021eba} and three flavor case~\cite{Chattopadhyay:2022ftv}. It may be noted that the treatment presented in~\cite{Chattopadhyay:2021eba, Chattopadhyay:2022ftv} did not incorporate the Majorana phases in the mixing matrix~\cite{ParticleDataGroup:2024cfk}.

In the present work, we use the most general $3 \times 3$ mixing matrix, which includes the two Majorana phases~\cite{ParticleDataGroup:2024cfk} and provide a complete analytic treatment (using Cayley-Hamilton formalism~\cite{Ohlsson:1999xb, Ohlsson:2001vp}) of neutrino oscillations in matter in the presence of decay. Also, contrary to the widespread belief, for the two-flavor case in vacuum, it was shown that Majorana phase could appear at the level of oscillation probability~\cite{Dixit:2022izn}. Here, for the three-flavor case and taking matter effects into account, we show that the two Majorana phases could become observable and appear at the level of oscillation probabilities. The off-diagonal terms (amplitude and phase) in the decay matrix play a crucial role in arriving at the above conclusion.  In the context of a long baseline neutrino experiment, we then analyse the impact of Majorana phases on the oscillation probabilities for different channels as well as on observables related to CP violation effects in neutrino oscillations.

This article is organised as follows. The formalism to calculate approximate neutrino oscillation probabilities in presence of neutrino decay is laid down  in Sec.~\ref{sec:formalism}.
Our results are described in Sec.~\ref{sec:results}. We compare the accuracy of our approximate analytical expressions with respect to numerical results in Sec.~\ref{sec:Approx_prob}.
In Sec.~\ref{sec:Majorana}, for the case of a long baseline experiment such as DUNE~\cite{Acciarri:2015uup, DUNE:2020ypp},  we show oscillation probabilities as a function of energy for the different cases (pure oscillation, oscillation with diagonal and general decay terms) considered in Sec.~\ref{sec:formalism}. 
In Sec.~\ref{sec:CPasym}, for the $\nu_{\mu}\to \nu_{e}$ channel, we explore the role of Majorana (and decay) phases on the  CP asymmetry. Finally, the effect of Majorana (and decay) phases on the parameter degeneracies in neutrino oscillations is described in   Sec.~\ref{sec:deg}. The concluding remarks are given in Sec.~\ref{sec:conclusion}.
In Appendix~\ref{sec:appA}, we briefly outline the Cayley-Hamilton formalism used for calculating the probabilities analytically. In Appendix~\ref{sec:appB}, we have included event level results for a benchmark case.

\section{Formalism}
\label{sec:formalism}

In the framework of neutrino oscillations with decay, likely, mass eigenstates and flavor eigenstates mismatch, and therefore a proper formalism to compute oscillation probabilities for the two flavor case was advocated in~\cite{Chattopadhyay:2021eba}.  For the three flavor case with $3 \times 3$ mixing matrix parameterized by three angles and one Dirac phase, analytic expressions were derived  in~\cite{Chattopadhyay:2022ftv,Gronroos:2024jbs}. Here, we additionally incorporate the two Majorana phases into the $3 \times 3$ mixing matrix and describe a complete analytic treatment for three-flavor neutrino oscillations coupled with decay. 
\subsection{Effective Hamiltonian for neutrino oscillations and  decay}
\label{sec:Ham_osc_dec}
In the presence of decay, the effective Hamiltonian no longer remains Hermitian. In general, we can express the effective Hamiltonian as
\bea
\label{eq:effective_H}
\mathcal{H} &=&\mathcal{H}_{\rm{h}} + \mathcal{H}_{\rm{ah}} = \mathcal{H}_{\rm{h}} 
-\dfrac{i}{2} \Gamma_{\rm{}}\,,
\eea \noindent
where $\mathcal{H}_{\rm h}$  is Hermitian while $\mathcal{H}_{\rm{ah}}= - i {\Gamma}/{2}$  is anti-Hermitian. The first term ($\mathcal{H}_{\rm h}$) describes neutrino oscillations in vacuum or in matter ($\mathcal{H}_{\rm h} = \mathcal{H}_{\rm v} + \mathcal{H}_{\rm m}$), while the second term ($\mathcal{H}_{\rm{ah}} = - i {\Gamma}/{2}$)  describes neutrino  decay. In the standard practice, the decaying states evolve according to the Wigner–Weisskopf approximation~\cite{Berryman:2014yoa, Chattopadhyay:2022ftv} where the probability of detecting the particle is not conserved. This leads to the corresponding effective Hamiltonian being non-Hermitian. 

Let us write down the explicit forms for $\mathcal{H}^{}_{h}$ and ${\Gamma^{}_{}}$ in the mass basis in which the $\mathcal{H}^{}_{h}$ is diagonal. In a vacuum (upon removing the term proportional to identity), we can write 
\bea
\label{eq:Ham_mass}
 \mathcal{H}  = 
 \begin{pmatrix}
 0 & 0 & 0\\
 0 & a^{}_{2} - a^{}_{1} & 0 \\
 0 & 0 & a^{}_{3} - a^{}_{1}
 \end{pmatrix} 
 - \dfrac{i}{2} 
 \begin{pmatrix}
2 d^{}_{1} & d^{}_{12}\text{e}^{i\chi^{}_{12}} & d^{}_{13}\text{e}^{i\chi_{13}}\\
d^{}_{12}\text{e}^{-i\chi_{12}} & 2 d^{}_{2} & d^{}_{23}\text{e}^{i\chi_{23}} \\
d^{}_{13}\text{e}^{-i\chi_{13}} & d^{}_{23}\text{e}^{-i\chi_{23}} & 2 d^{}_{3}
\end{pmatrix}\,,
\eea
\noindent
where  $a^{}_{i}$ ($ i=1,2,3$) can be identified as ${m^{2}_{i}}/({2E})$  with $m^{}_{i}$ being the mass of the $i$-{th} neutrino state and $E$ being the energy.  These are the eigenvalues of the effective Hamiltonian. 
The $d^{}_{i}, d^{}_{ij}$ for $i \neq j$ are the diagonal and off-diagonal decay parameters and $\chi^{}_{ij}$ are the corresponding phases of the off-diagonal decay terms. Note that $a^{}_{i}$, $d^{}_{i}$, $d^{}_{ij}$ and $\chi^{}_{ij}$ are real. 
Since decay is considered to be a sub-dominant effect,  it is understood that $ {\cal O}(a^{}_{i}) >  {\cal O} (d^{}_{ij}), {\cal O} (d^{}_{i})$. Furthermore, $\Gamma^{}_{}$ is restricted to be positive-definite.

In the three flavor case, the most general neutrino mixing matrix is parameterized by three angles $(\theta^{}_{12},\theta^{}_{23},\theta^{}_{13})$, the Dirac CP phase $(\delta^{}_{})$ and two Majorana phases $(\phi^{}_{1},\phi^{}_{2})$.
In the commonly adopted Pontecorvo-Maki-Nakagawa-Sakata (PMNS) parametrization~\cite{Pontecorvo:1957qd,Pontecorvo:1957cp, Maki:1962mu,Gribov:1968kq,ParticleDataGroup:2024cfk}, we have 
{\small \bea 
\label{eq:PMNS}
\mathcal{U} = 
\begin{pmatrix}
c^{}_{12}c^{}_{13} & s^{}_{12}c^{}_{13} & s^{}_{13} \text{e}^{-i\delta}\\
-s^{}_{12}c^{}_{23}-c^{}_{12}s^{}_{13}s^{}_{23} \text{e}^{i\delta} & c^{}_{12}c^{}_{23}-s^{}_{12}s^{}_{13}s^{}_{23} \text{e}^{i\delta} & c^{}_{13}s^{}_{23} \\
s^{}_{12}s^{}_{23}-c^{}_{12}s^{}_{13}c^{}_{23} \text{e}^{i\delta} & -c^{}_{12}s^{}_{23}-s^{}_{12}s^{}_{13}c^{}_{23} \text{e}^{i\delta^{}_{}} & c^{}_{13}c^{}_{23}
\end{pmatrix} 
\begin{pmatrix}
1 & & \\
& \text{e}^{i\phi_{1}}&
 \\
& & \text{e}^{i\phi_{2}}
\end{pmatrix} \,,
\eea }\noindent
where, $c^{}_{ij}=\cos\theta^{}_{ij}$ and $s^{}_{ij}=\sin\theta^{}_{ij}$. Using the form of $\mathcal{U}$ in Eq.~\eqref{eq:PMNS}, we can write down the effective matter Hamiltonian in the flavor basis 
{\small \bea
\label{eq:Gen_ham_dec}
\mathcal{H} &=& 
 \mathcal{U} \left[
  \frac{1}{2 E} 
  \begin{pmatrix} 
  	0& 0 & 0\\0 & \Delta m^{2}_{21} & 0\\0 & 0 & \Delta m^{2}_{31}
  \end{pmatrix} -  
  \dfrac{i}{{2}} \dfrac{\Delta m^{2}_{31}}{E} 
  \begin{pmatrix}
 \gamma^{}_{1} & \frac{1}{2}\gamma^{}_{12} \text{e}^{i\chi^{}_{12}} & \frac{1}{2}\gamma^{}_{13} \text{e}^{i\chi_{13}}\\
\frac{1}{2}\gamma^{}_{12} \text{e}^{-i\chi_{12}} &  \gamma^{}_{2} & \frac{1}{2}\gamma^{}_{23} \text{e}^{i\chi_{23}} \\
\frac{1}{2}\gamma^{}_{13} \text{e}^{-i\chi_{13}} & \frac{1}{2}\gamma^{}_{23} \text{e}^{-i\chi_{23}} &  \gamma^{}_{3}
  \end{pmatrix} 
 \right] \mathcal{U}^\dagger \nonumber 
 \eea
 \bea
\quad \quad \quad \quad  \quad \quad \quad \quad \quad \quad 
\quad \quad \quad \quad \quad \quad \quad \quad \quad \quad  \quad \quad \quad  \quad \quad \quad \quad
 &+&\begin{pmatrix} 
  	V^{}_{\rm{cc}} & 0 & 0\\0 & 0 & 0\\0 & 0 & 0
  \end{pmatrix}\,,
\eea }\noindent
where $\Delta m^{2}_{ij}=m^{2}_{i} - m^{2}_{j}$ is the mass-squared difference between $i$-th and $j$-th state. $V_{\rm{cc}}$ is the  matter potential induced due to the standard charged-current interactions of $\nu^{}_{e}$ with electrons, 
\bea
V^{}_{\rm{cc}}=\sqrt{2}\,G_{F} N_{e} = 7.6 \times 10^{-14}\,Y_{e} \,\left[\dfrac{\rho}{\rm{g/cc}}\right]\,\rm{eV}\,, \nonumber
\eea \noindent
where $G_F$ is the Fermi coupling constant and $N_e = N_{\rm{Avo}}\,\rho\, Y_{e}$ is the electron number density which depends on the mass density $(\rho)$, Avogadro's number, $N_{\rm{Avo}} = 6.023 \times 10^{23} \rm{g}^{-1} \rm{mol}^{-1}$ and the number of electrons per nucleon  ($Y_e \simeq 0.5$ for Earth matter). 
Here,  $\Delta m^2_{31} \gamma_{3} = m_{3}/\tau_{3}$, where $m^{}_{3}$ and $\tau^{}_{3}$ denote the mass and  lifetime respectively of the $\nu^{}_{3}$ state.
Usually, the Majorana phases are not observable at the level of oscillation probabilities if we only consider neutrino flavor oscillations~\cite{Bilenky:1980cx, Schechter:1980gr, Doi:1980yb, Giunti:2010ec}. However, for the scenario considered in the present work with the general decay matrix given by Eq.~\eqref{eq:Gen_ham_dec}, it turns out that the off-diagonal terms in the decay matrix can couple with the Majorana phases resulting in observable consequences at the level of oscillation probabilities. This was explicitly shown for the two flavor case in~\cite{Dixit:2022izn} and the condition for appearance of Majorana phase at the level of oscillation probabilities was derived. Here, we present a complete treatment of neutrino oscillations with decay using the general form of the $3 \times 3$ mixing matrix. We note that there are three distinct kinds of phases in Eq.~\eqref{eq:Gen_ham_dec} - (a)  Dirac CP phase ($\delta$), (b) Majorana phases ($\phi^{}_{1}$, $\phi^{}_{2}$), and (c) Decay phases $\chi^{}_{ij}\,(i,j = 1,2,3$ \textrm{and} $i \neq j)$ in the off-diagonal elements of the general decay matrix shown in Eq.~\eqref{eq:Gen_ham_dec}. We would like to remark that while Dirac CP phase appears at the level of detection probabilities in standard three flavor oscillation framework,  one can see imprints of the Majorana phases ($\phi^{}_{1}, \phi^{}_{2}$) at the level of detection probabilities if the off-diagonal decay terms are present.
In what follows, we compute the time-evolution operator using the Cayley-Hamilton formalism.
\subsection{Cayley-Hamilton formalism and the eigenvalues of the effective Hamiltonian}
\label{sec:eigen}
In order to compute the time-evolution operator, we use  Cayley-Hamilton formalism~\cite{Ohlsson:1999xb, Ohlsson:2001vp} for the effective Hamiltonian (Eq.~\eqref{eq:Gen_ham_dec}) described in Sec.~\ref{sec:Ham_osc_dec}. For details about the formalism, we refer the reader to Appendix~\ref{sec:appB}.
We are interested in obtaining the time-evolution operator
\bea
\label{eq:S_matrix}
{\mathcal S} &=& \text{e}^{{-i\mathcal{H} L}_{}} \, ,
\eea \noindent
where $\mathcal{H}$ is the effective Hamiltonian in the flavor basis (Eq.~\eqref{eq:Gen_ham_dec}) and $L$ is  the propagation distance. Using the Cayley-Hamilton formalism~\cite{
Ohlsson:1999xb, Ohlsson:2001vp} outlined in Appendix~\ref{sec:appB}, we obtain 
{\small \bea
    \label{eq:timeevolution}
\mathcal S = \text{e}^{-i{\mathcal{H} L}}_{} &=& \dfrac{\text{e}^{-i{ E_1}L}_{}}{(E^{}_1-E^{}_2)(E^{}_1-E^{}_3)}\Big[ \mathcal{H}^2_{} - (E^{}_2+E^{}_3)\mathcal{H} + E^{}_2 E^{}_3\boldsymbol{I} \Big] \nonumber \\ 
    &+&\dfrac{\text{e}^{-i E_2L}_{}}{(E^{}_2-E^{}_3)(E^{}_2-E^{}_1)}\Big[\mathcal{H}^2_{} - (E^{}_1 + E^{}_3)\mathcal{H} + E^{}_1 E^{}_3\boldsymbol{I} \Big] \nonumber \\
    &+&\dfrac{\text{e}^{-i E_3L}_{}}{(E^{}_3-E^{}_1)(E^{}_3-E^{}_2)}\Big[\mathcal{H}^2_{} - (E^{}_1+E^{}_2)\mathcal{H}  + E^{}_1 E^{}_2\boldsymbol{I} \Big]\,,
\eea }\noindent
where ${\mathcal S}$ is a complex $3\times 3$ matrix. $E^{}_{i}$ are the eigenvalues of the effective Hamiltonian (Eq.~\eqref{eq:Gen_ham_dec}). 
Once we have the time-evolution operator (${\mathcal S}$), we can immediately find the
 neutrino oscillation probabilities for different channels, $\nu^{}_\alpha \to \nu^{}_\beta$ $(\alpha$, $\beta = e$, $\mu$, $\tau)$ as $P^{}_{\alpha \beta} = \left|{\mathcal S}^{}_{\beta \alpha} \right|^2$.
 
We can express the energy eigenvalues $E^{}_{i}$ for the effective Hamiltonian in flavor basis (Eq.~\eqref{eq:Gen_ham_dec}) as a sum of the three contributions 
\bea
\label{eq:eigen_val}
E^{}_{i}&=&E^{(0)}_{i}+E^{(\gamma_{3})}_{i}+E^{(\Gamma)}_{i} \,,
\eea \noindent
where $i = 1, 2, 3$. Here, $E^{(0)}_{i}$ is  the $i$-th energy eigenvalue for the case when decay terms are absent, $E^{(\gamma_{3})}_{i}$  is  the $i$-th energy eigenvalue  due to diagonal decay case (with $\gamma^{}_{3}$ term only), $E^{(\Gamma)}_{i}$  is  the $i$-th energy eigenvalue for the general decay scenario. 
We note that $\alpha, s^{}_{13}, \gamma^{}_{3}$ are small parameters,
\bea
\label{eq:exact1}
\alpha&=&\frac{\Delta m^{2}_{21}}{\Delta m^{2}_{31}} \simeq 0.03,~~~~~~~~ s^{}_{13} \simeq 0.14,~~~~~~~\gamma^{}_{3} \lesssim 0.1\,.
\eea
Given their values, we can express them in terms of powers of $\lambda$ as 
\bea
\label{eq:approx1}
\alpha &\simeq& 0.03 \simeq \mathcal{O}(\lambda^{2}_{}),~~~~~~~~ s^{}_{13} \simeq 0.14 \simeq \mathcal{O}(\lambda),~~~~~~~\gamma^{}_{3} \lesssim 0.1  \simeq \mathcal{O}(\lambda) \,,
\eea \noindent
where $\lambda \equiv 0.2$ is simply a book-keeping parameter. In certain flavour-symmetric models $\lambda$ can be connected to the Wolfenstein parameter in  the CKM sector. And, there have been attempts to connect such a parameter to neutrino sector as well~\cite{Boucenna:2012xb,Xing:2024pal}. Here we remain agnostic about that and choose this typical value which 
is consistent with the latest neutrino oscillation data.
We also note that $\gamma^{}_{1} \Delta m^2_{31} \lesssim {\mathcal O} (\lambda) \Delta m^2_{21}$,   $\gamma^{}_2 \Delta m^2_{31} \lesssim {\mathcal O} (\lambda) \Delta m^2_{21}$ and  $\gamma^{}_3 \Delta m^2_{31} \lesssim {\mathcal O} (\lambda) \Delta m^2_{31}$~\cite{Chattopadhyay:2022ftv}. By taking $\Gamma$ to be positive-definite, the off-diagonal terms are constrained to be   
${\mathcal O} (\gamma^{2}_{ij}) \lesssim {\mathcal O} (\gamma^{}_{i}) {\mathcal O} (\gamma^{}_{j})$. Thus, 
\bea
\label{eq:approx2}
	\gamma^{}_{1}\,,\, \gamma^{}_{2}\sim \mathcal{O}(\lambda^3_{})\,,\qquad \gamma^{}_{3} \sim \mathcal{O}(\lambda)\,,\qquad \gamma^{}_{12} \sim \mathcal{O}(\lambda^3_{})\,,\qquad \gamma^{}_{13}\,,\,\gamma^{}_{23} \sim \mathcal{O}(\lambda^2_{})\,.
\eea
Note that in  deriving the energy eigenvalues and the oscillation probabilities, we apply the approximations in Eqs.~\eqref{eq:approx1} and \eqref{eq:approx2}, with all expressions organized as expansions in powers of the book-keeping parameter $\lambda$.
The approximate form  of $E^{(0)}_{i}$ is given by
\bea
\label{eq:osc}
		E^{(0)}_{1}&=&\dfrac{\Delta m^{2}_{31}}{2 E}\Big\{A+\alpha_{}^{}  s^{2}_{12}+ s^{2}_{13} \dfrac{A }{A-1}\Big\}+{\mathcal{O}}(\lambda^3_{})\,,  \\
		E^{(0)}_{2}&=&  \dfrac{\Delta m^{2}_{31}}{2 E}\big\{\alpha_{}^{} c^{2}_{12}\big\}+{\mathcal{O}}(\lambda^3_{})\,,
		\\
		E^{(0)}_{3}&=& \dfrac{\Delta m^{2}_{31}}{2 E} \Big\{1- s^{2}_{13} \dfrac{A}{A-1}\Big\}+{\mathcal{O}}(\lambda^3_{})\,,
\eea \noindent
where, $A = {2 V_{\rm{cc}} E}/({\Delta m_{31}^2})$ is a dimensionless parameter. Note that we have retained terms up to ${\mathcal{O}}(\lambda^2)$ here. 
The approximate form  of $E^{(\gamma^{}_{3})}_i$ is given by
\bea 
\label{eq:gamma3}
E^{(\gamma_3)}_{1}&=& \dfrac{\Delta m^{2}_{31}}{2 E}\left\{-i \gamma^{}_{3} s^{2}_{13} \dfrac{A^2}{(A-1)^2}\right\}+{\mathcal{O}}(\lambda^4_{}) \,, \\
		E^{(\gamma_3)}_{2} &=& 0  \,, \\ 
		E^{(\gamma_3)}_{3} &=& \dfrac{\Delta m^{2}_{31}}{2 E} \left\{-i \gamma^{}_{3}+i \gamma^{}_3 s^{2}_{13} \dfrac{ A^2}{(A-1)^2}\right\}+{\mathcal{O}}(\lambda^4).
\eea
These expressions are consistent with those obtained in~\cite{Chattopadhyay:2022ftv,Gronroos:2024jbs}. Note that we have retained terms up to ${\mathcal{O}}(\lambda^3_{})$ here.  
$E^{}_{1}$ gets contribution from $\gamma^{}_{3}$ at ${\mathcal{O}}(\lambda^3_{})$, $E^{}_{2}$ does not get any modification to this order and $E^{}_{3}$ has contribution at ${\mathcal{O}}(\lambda^{}_{})$. It may be noted that the Majorana phases $(\phi^{}_{1}, \phi^{}_{2})$ do not appear when we consider diagonal decay matrix with $\gamma^{}_{3}$ non-zero (see Eq.~\eqref{eq:gamma3}). 
Proceeding in the same way, we can express the approximate form  of $E^{(\Gamma)}_{i}$ as
{\small \bea 
\label{eq:Gamma}
E^{(\Gamma)}_{1}&=& \dfrac{\Delta m^{2}_{31}}{2 E} \Big\{-i \gamma^{}_{1} c^{2}_{12}-i \gamma^{}_{2}  s^{2}_{12}-i \gamma^{}_{12}  s^{}_{12} c^{}_{12} \cos [\phi^{}_{1}-\chi^{}_{12}] 
-i s^{}_{13} \Big(\gamma^{}_{13}  c^{}_{12}\cos \left[\delta^{}_{}-\phi^{}_{2}+\chi ^{}_{13}\right]\nonumber \\&+&\gamma^{}_{23}  s^{}_{12} \cos \left[\delta^{}_{}+\phi^{}_{1}-\phi^{}_{2}+\chi^{}_{23}\right]\Big) \dfrac{A}{A-1} \Big\}+{\mathcal{O}}(\lambda^4_{}) \,,\\ 
E^{(\Gamma)}_{2}&=&\dfrac{\Delta m^{2}_{31} }{2 E}\Big\{-i \gamma^{}_{1} s^{2}_{12}-i\gamma^{}_{2}  c^{2}_{12}-i \gamma^{}_{12}  s^{}_{12} c^{}_{12}  \cos [\phi^{}_{1}-\chi^{}_{12}] \Big\}+{\mathcal{O}}(\lambda^4_{})\,, \\
E^{(\Gamma)}_{3}&=&\dfrac{\Delta m^{2}_{31}}{2 E}\Big\{i  s^{}_{13} \Big(\gamma^{}_{13}  c^{}_{12} \cos \left[\delta^{}_{}-\phi^{}_{2}+\chi^{}_{13}\right]+\gamma^{}_{23}\, s^{}_{12} \cos \left[\delta^{}_{}+ \phi^{}_{1}-\phi^{}_{2}+\chi^{}_{23}\right]\Big)\dfrac{A}{A-1}\Big\} \nonumber \\ & + & {\mathcal{O}}(\lambda^4)\,.
\eea } \noindent
Note that here also we have retained terms up to ${\mathcal{O}}(\lambda^3_{})$. Unlike the previous case, here $E^{}_{1}, E^{}_{2} $, and $ E^{}_{3}$ would have contribution from the diagonal and off-diagonal elements of $\Gamma$ at ${\mathcal{O}}(\lambda^3_{})$.

A careful examination of the approximate eigenvalues reveals that the Majorana phases $(\phi_{1}, \phi_{2})$ appear in the expression for general decay scenario, governed by $\Gamma$ (see Eq.~\eqref{eq:Gamma}).
The Majorana phases arise due to the off-diagonal terms $\gamma^{}_{ij}\,(i,j = 1,2,3$ \textrm{and} $i \neq j)$ in the decay matrix and it may also be noted that the Dirac phase $(\delta)$, Majorana phase $(\phi^{}_{1},\phi^{}_{2})$ and decay phases $\chi^{}_{ij}\,(i,j = 1,2,3$ \textrm{and} $i \neq j)$ appear together. This implies that the three sources of CP violation appear in an entangled way. These three sources may interfere constructively or destructively depending on the channel and values of these phases. One can examine the precise role of a given source or phase term by setting all except that one parameter non-zero. 

\subsection{Approximate probability expressions}
\label{sec:prob}
We can express the probability as a sum of the three contributions 
\bea
P^{}_{\alpha \beta} &=&P^{(0)}_{\alpha \beta}  
+ P^{(\gamma_3)}_{\alpha \beta}+P^{(\Gamma)}_{\alpha \beta}  \,\nonumber.
\eea \noindent
In what follows, we write down the approximate expressions for different channels. 
For the $\nu^{}_{\mu} \to\nu^{}_{e}$ channel, we have 
{\small \bea
\label{eq:prob_me}
P^{(0)}_{\mu e} &=& 4 s^{2}_{13} s^{2}_{23} \dfrac{\sin ^2[(A-1) \Delta]}{(A-1)^2}
 + 2 \alpha^{}_{}  s^{}_{13} \sin 2 \theta^{}_{12} \sin 2 \theta^{}_{23} \cos \left(\Delta^{}_{} +\delta^{}_{}\right)\dfrac{\sin [(A-1) \Delta ]}{A-1}\dfrac{\sin A \Delta}{A}\nonumber \\&+&{\mathcal{O}}(\lambda^4_{})\, , 
 \\[1mm]
P^{(\gamma_{3})}_{\mu e}&=& -8 \gamma^{}_{3}  s^{2}_{13} s^{2}_{23}  \Delta^{}_{}  \dfrac{\sin ^2[(A-1) \Delta]}{(A-1)^2}+{\mathcal{O}}(\lambda^4_{})\,,
 \\[1mm] 
P^{(\Gamma)}_{\mu e} &= & 4 s^{}_{13} s^{2}_{23} \Big\{ \gamma^{}_{13} c^{}_{12} \sin [{\delta^{}_{} - \phi^{}_{2}+\chi^{}_{13}}]+\gamma^{}_{23} s_{12} \sin [{\delta^{}_{} + \phi^{}_{1}- \phi^{}_{2}+\chi^{}_{23}}]\Big\}\dfrac{\sin^{2}[ (A-1) \Delta]}{(A-1)^{2}}\nonumber \\&+& {\mathcal{O}} (\lambda^{4}_{})\,,
\eea }\noindent
where $\Delta = \Delta m^{2}_{31} L/4E$ is a dimensionless parameter.
For the  $\nu^{}_{e} \to\nu^{}_{e}$ channel, we have
{\small \bea
\label{eq:prob_ee}
P^{(0)}_{ee}&=&1-4 s^{2}_{13} \dfrac{\sin ^2 [(A-1)\Delta]}{(A-1)^2}+{\mathcal{O}}(\lambda^4) \,, \\[1mm]
 	P^{(\gamma_3)}_{ee}&=&\gamma^{}_3  s^{2}_{13} \left\{4 A\dfrac{\sin [2 (A-1) \Delta]}{(A-1)^3}-4 \Delta \dfrac{1+ A^2}{(A-1)^2}+8\Delta^{}_{} \dfrac{\sin^2 [(A-1)\Delta]}{(A-1)^2}\right\}+{\mathcal{O}}(\lambda^4)\,,\\[1mm]
	P^{(\Gamma)}_{ee}&=&-4 \gamma^{}_{1}  c^{2}_{12} \Delta^{}_{} -4 \gamma^{}_{2} s^{2}_{12} \Delta^{}_{}  -2 \gamma^{}_{12} \Delta^{}_{}  \sin 2 \theta^{}_{12} \cos [\phi^{}_{1} -\chi^{}_{12}]+ 2 s^{}_{13} \Big\{\gamma^{}_{13}  c^{}_{12} \cos \left[\delta_{}-\phi^{}_{2} +\chi^{}_{13}\right] \nonumber \\&+&\gamma^{}_{23}  \, s^{}_{12} \cos \left[\delta^{}_{}+\phi^{}_{1} -\phi^{}_{2} +\chi^{}_{23}\right]\Big\} \left\{\dfrac{\sin [2 (A-1) \Delta] }{(A-1)^2}-\dfrac{2 A \Delta }{A-1}\right\} +{\mathcal{O}}(\lambda^4)\, .
\eea }\noindent
We see that for $\nu^{}_{\mu}\to \nu^{}_{e}$ and $\nu^{}_{e}\to \nu^{}_{e}$ channels, the Majorana phase dependence is captured via terms of $\mathcal{O}(\lambda^{3}_{})$.  
For the $\nu^{}_{\mu} \to\nu^{}_{\mu}$ channel, 
{\small \bea
\label{eq:prob_mm}
P^{(0)}_{\mu\mu} &=& 1-\sin ^{2} 2 \theta_{23} \sin ^2\Delta_{} 
- \dfrac{2}{A-1} s^{2}_{13}  \sin^{2} 2\theta_{23} \Big\{\sin\Delta \cos A\Delta  \dfrac{\sin[(A-1) \Delta]}{A-1}-\dfrac{A}{2}\Delta \sin 2\Delta\Big\}\nonumber\\
&-& 4 s^{2}_{13} s^{2}_{23} \dfrac{\sin^{2}[(A-1) \Delta ]}{(A-1)^2}+\alpha^{}_{} \, c^{2}_{12}  \sin^{2} 2\theta_{23}  \Delta \sin 2\Delta +{\mathcal{O}}(\lambda^3_{})\,, \\[1mm] 
P^{(\gamma_3)}_{\mu\mu}&=& -\gamma^{}_{3} \Delta^{}_{}  \left(\sin ^{2} 2 \theta_{23}\cos 2 \Delta +4 s^{4}_{23}\right) + \gamma^{2}_{3} \Delta^2_{} \left(\sin^{2} 2\theta^{}_{23} \cos 2 \Delta +8 s^{4}_{23}\right)+{\mathcal{O}}(\lambda^3_{})\,,  \\[1mm]
P^{(\Gamma)}_{\mu\mu}&=&\sin 2 \theta^{}_{23} \Big\{\gamma^{}_{13} \, s^{}_{12} \cos [\phi^{}_2 -\chi^{}_{13}] -\gamma^{}_{23}  c^{}_{12} \cos [\phi^{}_{1} -\phi^{}_{2} +\chi^{}_{23}]\Big\}\sin 2 \Delta +{\mathcal{O}}(\lambda^3)\,.
\eea }\noindent
We see that for $\nu^{}_{\mu}\to \nu^{}_{\mu}$  channel, the Majorana phase dependence is captured via terms of $\mathcal{O}(\lambda^{2}_{})$.  
In the limit of vanishing Majorana phases ($\phi^{}_{1}\simeq \phi^{}_{2} \simeq 0$),  our analytical results (Eqs.~\eqref{eq:prob_me} - \eqref{eq:prob_mm}) are consistent with~\cite{Chattopadhyay:2022ftv,Gronroos:2024jbs}. 
The anti-neutrino oscillation probabilities can be readily obtained by noting,
	\bea
		P_{\bar{\alpha}\bar{\beta}}^{}&=&P_{\alpha\beta} (\delta^{}_{} \to -\delta^{}_{}\, ,\phi^{}_{i} \to -\phi^{}_{i} \, ,\, \chi^{}_{ij} \to -\chi^{}_{ij}\, , \, A \to -A)\,.
		\label{eq:cpasym}
	\eea
It is worth pointing out that  the influence of the decay parameters is small,  i.e., arises at $\mathcal{O}(\lambda^{2}_{})$ or $\mathcal{O}(\lambda^{3}_{})$, in order to assess the impact, one would require a high-precision experiment such as DUNE~\cite{Acciarri:2015uup, DUNE:2020ypp}.

 \begin{table}[t]
\centering
\scalebox{0.8}{
\begin{tabular}{| c | c | c | c |}
\hline
&&&\\
Experiments & $ \tau^{}_{3}/m^{}_{3}$ & $\gamma^{}_{3}$ 
& References \\[2mm]
\hline \hline &&&\\
T2K + NO$\nu$A  & $1.5 \times 10^{-12}$\,(3\,$\sigma$)    &  0.174  & \cite{Choubey:2018cfz}\\[2mm]
T2K + NO$\nu$A + MINOS &  $2.4 \times 10^{-11}$\,($90\% \,$C.L.)                  &  0.0109  & \cite{Ternes:2024qui} \\[2mm]
SK + MINOS &  $2.6 \times 10^{-12}$\,($99\%$\,C.L.)               & 0.10 & \cite{Gonzalez-Garcia:2008mgl} \\[2mm] 
\hline &&& \\
JUNO &  $5.5 \times 10^{-11}$\,($99\%$\,C.L.)                 &  0.00476& \cite{Abrahao:2015rba} \\[2mm]
ESS$\nu$SB  & $2.43 \times 10^{-11}$\,(3\,$\sigma$)  &  0.0107 & \cite{Chakraborty:2020cfu} \\[2mm]
ESS$\nu$SB + T2HK  & $4.36 \times 10^{-11}$\,(3\,$\sigma$)  &  0.00601 & \cite{Chakraborty:2020cfu} \\[2mm]
DUNE  & $4.22 \times 10^{-11}$\,(3\,$\sigma$) &  0.00621 & \cite{Dey:2024nzm} \\[2mm]
P2SO  & $2.11 \times 10^{-11}$\,(3\,$\sigma$) &  0.0124 & \cite{Panda:2024ioo} \\[2mm]
&&&\\
\hline
\end{tabular}}
\vskip 0.1in
\caption{\label{tab:decay_parameters}
\footnotesize $\tau^{}_{3}/m^{}_{3}$ and $\gamma_3$ bounds obtained from different experiments. The constraints for SK+MINOS, T2K+NO$\nu$A, and T2K+NO$\nu$A+MINOS are derived from experimental data.}
\end{table}

\section{Results}
\label{sec:results}

In what follows, we shall use the standard three flavor oscillation parameters consistent with a global analysis of neutrino oscillation data~\cite{deSalas:2020pgw} (see Table~\ref{tab:parameters}). 
The value of the Dirac CP phase is taken to be $\delta = -\pi/2$ as indicated by the global analyses for NH case~\cite{deSalas:2020pgw} (see also \cite{Capozzi:2017ipn, Esteban:2024eli}), however $ \delta^{}_{}$ could take any value, $ \delta \in [-\pi,\pi]$.
The two Majorana phases could take any value,  $ \phi^{}_{i} \in [-\pi,\pi]~(i=1,2)$.
Concerning the decay parameters, we take values consistent with the criteria given by Eq.~\eqref{eq:approx2}. We take the following  values for the decay parameters: 
\bea
\gamma^{}_{1} &=& 0.001\,, \gamma^{}_{2} = 0.001\,, \gamma^{}_{3} = 0.1\,, \gamma^{}_{12} = 0.001\,, \gamma^{}_{13} = 0.01\,, \gamma^{}_{23} = 0.01 \,.
\label{eq:para_decay}
\eea
In Table~\ref{tab:decay_parameters}, we summarize the latest as well as prospective bounds on $\gamma_3$ from various experiments.  It is to be noted that various experiments put bounds on $m^{}_{3}/\tau^{}_{3}$.     By using $\gamma_3 \times\ldm=m^{}_{3}/\tau^{}_{3}$, we can obtain the corresponding bounds $\gamma_3$.  In our analysis, we adopt the value of $\gamma_3$ which is consistent with the combined limit arising from T2K + NO$\nu$A~\cite{Choubey:2018cfz}. For the remaining $\gamma_i$'s, we follow the parametrization given by Eq.~\eqref{eq:approx2}.
The decay phases can take any value within the allowed range,  $ \chi^{}_{ij} \in [-\pi,\pi]~(i,j=1,2,3 ~ \textrm{and} ~ i \neq j)$. Note that the primary goal of the present work is to bring out the effect of Majorana phases at  the level of probabilities so the discussion is centered around probabilities. For the sake of completeness, we have also included event level results for a benchmark case in Appendix~\ref{sec:appB}. 

 \begin{table}[t]
\centering
\scalebox{0.8}{
\begin{tabular}{| c | c | c | c |}
\hline
&&&\\
Parameter & Best-fit-value & 3$\sigma$ interval   & $1\,\sigma$ uncertainty\\[2mm]
\hline \hline &&&\\[-2mm]
$\theta^{}_{12}/^\circ$            & 33.68                   &  31.63 - 35.95 &  2.9\% \\[2mm]
$\theta^{}_{13}/^\circ$ (NH)    & 8.56            &  8.19  -  8.89 &  1.5\%  \\ [2mm]
$\theta^{}_{13}/^\circ$ (IH)     & 8.59              &  8.25  -  8.93 & 1.5\%\\[2mm]
$\theta^{}_{23}/^\circ$ (NH)        & 43.3          &  41.3  - 49.9 & 3.5\%\\ [2mm]
$\theta^{}_{23}/^\circ$ (IH)        & 47.9          &  41.5  - 49.8 & 3.5\%\\ [2mm]
$\delta^{}_{}/^\circ$ (NH)   & 212   & 124 - 364 & - \\[2mm]
$\delta^{}_{}/^\circ$ (IH)   & 274   & 201 - 335   & - \\[2mm]
$\sdm$ [$\text{eV}^2$]  & 7.49 $\times 10^{-5}$  &  [6.92 - 8.05] $\times 10^{-5}$ & 2.7\%  \\[2mm]
$\ldm$ (NH) [$\text{eV}^2$] & $+$2.51 $\times 10^{-3}$   &  [2.45 - 2.57] $\times 10^{-3}$ & 1.2\% \\[2mm]
$\ldm$ (IH) [$\text{eV}^2$] & $-$2.48 $\times 10^{-3}$  & $-$[2.54 - 2.42] $\times 10^{-3}$  & 1.2\% \\
&&&\\
\hline
\end{tabular}}
\vskip 0.1 in
\caption{\label{tab:parameters}
\footnotesize Best-fit values and $1\,\sigma$ uncertainties of the three flavor neutrino oscillation parameters and their $3\,\sigma$ range taken from~\cite{Esteban:2024eli} (see also \cite{Capozzi:2017ipn, deSalas:2020pgw}). Here, NH (IH) refers to the normal (inverted) neutrino mass hierarchy. 
 }
\end{table}
\subsection{Accuracy of the approximate series expansions of the probabilities}
\label{sec:Approx_prob}
The analytic probability expressions (Eqs.~\eqref{eq:prob_me} -\eqref{eq:prob_mm}) for the relevant channels are  obtained in Sec.~\ref{sec:prob} are only approximate. Let us  first assess the validity and accuracy of the series expansions with respect to the numerical results. 
For the purpose of quantitative illustration, we consider  $L=1300$ km and average density $\rho = 2.848$ g/cc which correspond to the upcoming long baseline experiment, DUNE~\cite{Acciarri:2015uup, DUNE:2020ypp}. We would like to remark that the conclusions drawn would be valid for any other current and future neutrino long baseline experiment, such as T2K~\cite{T2K:2017hed}, T2HK/T2HKK~\cite{Hyper-KamiokandeProto-:2015xww, Hyper-Kamiokande:2016srs, Hyper-Kamiokande:2018ofw} and NO$\nu$A~\cite{NOvA:2016kwd}. 
We use the parameter values given in Table~\ref{tab:parameters} (for standard parameters) and Eq.~\eqref{eq:para_decay} (for decay parameters). The values of the Dirac phase, Majorana phases and decay phases taken are indicated on the plots.

In our approximate expressions, we took $\alpha, s^{}_{13} \ll 1$, which means that we can identify the regime of validity of our expressions. The expressions would show poor validity once the oscillations due to the solar mass-squared splitting become relevant, i.e., when 
\bea
\alpha \dfrac{\Delta m^2_{31} L}{4 E}  &\sim& {\mathcal O} (1)\,. \nonumber
\eea
This happens at $L/E \sim 10^{4}$ km/GeV. 
Thus, our expressions will be valid when $L/E \ll  10^4$ km/GeV. In other words, it is expected that our expansion becomes imprecise for very long baselines or at very low energies.

\begin{figure}[t!]
\hskip -0.3in
\centering
\includegraphics[width=6.8in]{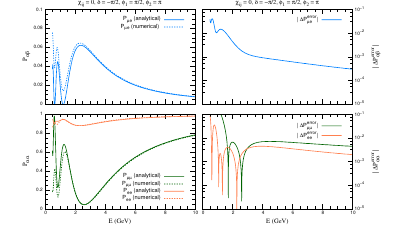}
\vspace{-1cm}
\caption{\footnotesize Left panel : Comparison between analytical expression and numerical result for appearance (top) and disappearance (bottom) channels for the case of DUNE ($L = 1300$ km). The solid (dotted) line represents analytical expressions (numerical results). Right panel : the absolute error, $|\Delta P^{\textrm{error}}_{\alpha \beta}|$ (top) and $|\Delta P^{\textrm{error}}_{\alpha \alpha}|$ (bottom).}
\label{fig:prob_error}
\end{figure}

In Fig~\ref{fig:prob_error},  we compare analytical expressions  and numerical predictions for different channels, $\nu^{}_{\mu}\to \nu^{}_{e}$, $\nu^{}_{\mu}\to \nu^{}_{\mu}$ and $\nu^{}_{e}\to \nu^{}_{e}$ for the representative case of DUNE. We quantify the precision of our expressions in terms of absolute error between the analytical expression and the numerically obtained probabilities, 
\bea
\label{eq:prob_diff1}
|\Delta P^{\textrm{error}}_{\alpha \beta}| &=& |P^{(\rm{analytical})}_{\alpha \beta}- P^{(\rm{numerical})}_{\alpha \beta}|\,.
\eea
The left (right) panel depicts the oscillation probability (absolute error between analytical and numerical prediction) for the given channels.  
If we consider the $\nu^{}_{\mu}\to \nu^{}_{e}$ channel (top row in Fig~\ref{fig:prob_error}), the two estimates are in fair agreement (i.e., $|\Delta P^{\textrm{error}}_{\mu e}|  < 0.001$) at energies beyond $2$ GeV. In particular, in the region near the peak of $\nu^{}_{\mu}\to \nu^{}_{e}$ probability (for $L = 1300$ km, $E=2.3$ GeV), the analytical expression works well. 
For the $\nu^{}_{\mu}\to \nu^{}_{\mu}$ channel (bottom row in Fig~\ref{fig:prob_error}), the two estimates agree well (i.e., $|\Delta P^{\textrm{error}}_{\mu \mu}|  < 0.001$) at energies beyond $1.8$ GeV. 
For the $\nu^{}_{e}\to \nu^{}_{e}$ channel (bottom row in Fig~\ref{fig:prob_error}), the two estimates agree well (i.e., $|\Delta P^{\textrm{error}}_{ee}|  < 0.001$) at energies beyond $1.1$ GeV.  Generally speaking, the agreement is fair for energies beyond $\gtrsim 1.5$ GeV (depending on the channel). In general this is true for other long baseline experiments too. 
We have checked that the qualitative conclusion regarding accuracy of the analytical expressions is valid even for the choice of phases (in the allowed range) other than those used for illustrative purposes ($\delta^{}_{} = -\pi/2$, $\phi^{}_{1} = \pi/2, \phi^{}_{2} = \pi$, $\chi^{}_{ij} =0$). 
\subsection{Impact of Majorana phases and decay phases on the oscillation framework}
\label{sec:Majorana}
\begin{figure}[t!]
\hskip -0.3in
\includegraphics[width=7.5in]{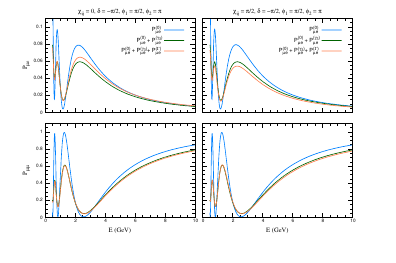}
\vspace{-1.8cm}
\caption{\footnotesize The upper (lower) panel corresponds to $P^{}_{\mu e}$ ($P^{}_{\mu \mu}$) as a function of energy $E$. Different curves are shown for no decay, $P^{(0)}_{\mu e} $, no decay together with only $\gamma^{}_3$ decay, $P^{(0)}_{\mu e} +P^{(\gamma_{3})}_{\mu e}$, and for the most general case,  $P^{(0)}_{\mu e} +P^{(\gamma_{3})}_{\mu e}+ P^{(\Gamma)}_{\mu e}$, using the blue, orange, and green lines, respectively, for DUNE. The left and right panels differ due to the benchmark values of the phases, as mentioned at the top of the plots.
}
\label{fig:prob_num_matt}
\end{figure}

The discussion of three flavor neutrino oscillation probabilities in presence of neutrino decay (see Sec.~\ref{sec:formalism}) reveals that non-zero off-diagonal elements $\gamma^{}_{ij}$\,$(i,j=1,2,3$ \textrm{and} $i \neq j$) in the decay matrix are responsible for the appearance of  the Majorana phase term at the level of probabilities (see Eqs.~\eqref{eq:prob_me} -\eqref{eq:prob_mm} for analytical expression of probabilities for different channels). Since  decay  appears as a sub-dominant effect over neutrino oscillations, and we would like to assess the role of off-diagonal elements in the decay matrix, we follow a numerical approach to capture the effects precisely. We shall see later that the analytical expressions 
 allow for a clear  understanding of the dependence on different parameters.
Fig.~\ref{fig:prob_num_matt} depicts  $\nu^{}_{\mu} \to \nu^{}_{e}$ probability (top panel)  and   $\nu^{}_{\mu} \to \nu^{}_{\mu}$  probability (bottom panel). The plots on the left are for $\chi^{}_{i j} = 0$ $(i,j = 1,2,3$ \textrm{and} $i \neq j)$ while non-zero decay phases $\chi^{}_{i j} = {\pi}/{2}$ $(i,j = 1,2,3$ \textrm{and} $i \neq j)$ are considered in the right panel.
Following the discussion in Sec.~\ref{sec:formalism}
(see Eqs.~\eqref{eq:prob_me} -\eqref{eq:prob_mm}), we consider three cases of Fig.~\ref{fig:prob_num_matt} as 
\begin{enumerate}
\item
[(a)] neutrino oscillation in matter and no decay, shown in blue, 
\item
[(b)] neutrino oscillation in matter with diagonal decay term ($\gamma^{}_{3}$), shown in green, and 
\item[(c)] neutrino oscillation in matter with general decay term ($\Gamma^{}_{}$), shown in orange. 
\end{enumerate}
In what follows, we shall discuss how diagonal and off-diagonal decay terms influence the oscillation signals which may be captured by long baseline oscillation experiments. 
In order to understand the dependence of Majorana phases $\phi^{}_{i}\,(i=1,2)$ and the interplay of the three different phases, we note that (see Eq.~\eqref{eq:prob_me})
\bea
P^{(\gamma_{3})}_{\mu e} &\propto&  -8 \gamma^{}_{3}  s^{2}_{13} s^{2}_{23} \,
\eea
\bea
\label{eq:MajoranaPhase_mue}
P^{(\Gamma)}_{\mu e} &\propto&   \gamma^{}_{13} c^{}_{12} \sin [{\delta^{}_{} - \phi^{}_{2}+\chi^{}_{13}}] + \gamma^{}_{23} s^{}_{12} \sin [{\delta^{}_{} + \phi^{}_{1}- \phi^{}_{2}+\chi^{}_{23}}]\,.
\eea
which implies that non-zero  $\gamma^{}_{ij}$\,$(i,j=1,2,3$ \textrm{and} $i \neq j$)  are solely responsible for seeing the effect of Majorana phases at the probability level. We also note that $\chi^{}_{ij} $ $(i,j = 1,2,3$ \textrm{and} $i \neq j)$ appear in coupled way along with  other phase terms $\delta$ and $\phi^{}_{i}\,(i=1,2)$. 
For the left panel, we have taken  $\chi^{}_{i j} = 0$, $\delta^{}_{} = -\pi/2, \phi^{}_{1} = \pi/2,  \phi^{}_{2} = \pi$, which leads to  $P_{\mu e}^{(\Gamma)} \propto  \gamma^{}_{13}$. For the right panel, we have taken $\chi^{}_{i j} = \pi/2$, $\delta^{}_{} = -\pi/2, \phi^{}_{1} = \pi/2,  \phi^{}_{2} = \pi$, which leads to  $P_{\mu e}^{(\Gamma)} \propto - \gamma^{}_{23}$. 
%
From Fig.~\ref{fig:prob_num_matt} (top panel), we  note that in the region  around $E \simeq 2.3$ GeV (first oscillation maximum), the three cases ($(a)$, $(b)$, $(c)$) are clearly distinct.  For case $(b)$,  $P^{(\gamma_{3})}_{\mu e} \propto -  \gamma^{}_{3} $ and given the choice of decay parameters (Eq.~\eqref{eq:para_decay}), this term dominates over the other terms in Eq.~\eqref{eq:prob_me}.  Thus, the overall probability with the general form of the decay matrix  is in general lower in comparison to case $(a)$. 
In case (c),  $P^{(\Gamma_{})}_{\mu e}$ will have contribution from $\gamma^{}_{3}$ as well as $\gamma^{}_{13}$ with opposite signs for the choice of parameters in left panel. 
For the right panel,  since $\chi^{}_{i j} = \pi/2$,  $P^{(\Gamma)}_{\mu e}$ will have contribution from $\gamma^{}_{3}$ as well as $\gamma^{}_{23}$ (with same sign), and the curves corresponding to $(b)$ and $(c)$ get interchanged. 

From the bottom panel, we note that the $\nu^{}_{\mu} \to \nu^{}_{\mu}$ channel is also impacted  due to neutrino decay terms. However, for the general decay parameters considered,  case (b) and (c) show similar behaviour. 
In order to understand the dependence of Majorana phases $\phi^{}_{i}$\,($i=1,2$) and the interplay of the three different phases, we note that (see Eq.~\eqref{eq:prob_mm})
\bea
P^{(\gamma_3)}_{\mu\mu}&\propto& -\gamma^{}_{3} \Delta^{}_{}  \left(\sin ^{2} 2 \theta_{23}\cos 2 \Delta +4 s^{4}_{23}\right) + \gamma^{2}_{3} \Delta^2_{} \left(\sin^{2} 2\theta^{}_{23} \cos 2 \Delta +8 s^{4}_{23}\right)\,,
\eea
\bea
\label{eq:MajoranaPhase_mumu}
P^{(\Gamma)}_{\mu \mu} &\propto&   \gamma^{}_{13} s^{}_{12} \cos [{\phi^{}_{2}-\chi^{}_{13}}] - \gamma^{}_{23} c^{}_{12} \cos [{\phi^{}_{1}- \phi^{}_{2}+\chi^{}_{23}}] \,.
\eea
For the left panel, we have taken $\chi^{}_{i j} = 0$, $\delta^{}_{} = -\pi/2, \phi^{}_{1} = \pi/2,  \phi^{}_{2} = \pi$, which leads to  $P^{(\Gamma)}_{\mu \mu} \propto -\gamma^{}_{13}$. For the right panel, we have taken $\chi^{}_{i j} = \pi/2$, $\delta^{}_{} = -\pi/2, \phi^{}_{1} = \pi/2,  \phi^{}_{2} = \pi$, which leads to  $P^{(\Gamma)}_{\mu \mu} \propto -\gamma^{}_{23}$. 
$P^{(\Gamma)}_{\mu \mu} $ will have contribution from $\gamma^{}_{3}$ as well as $\gamma^{}_{13}$ (with same sign) for left plot. $P^{(\Gamma)}_{\mu \mu}$ will have contribution from $\gamma^{}_{3}$ as well as $\gamma^{}_{23}$ (with same sign) for the right plot. One can note that $\delta$ does not appear in Eq.~\eqref{eq:prob_mm}. 
We note that contribution from the off-diagonal elements in the general decay matrix is very small (curves for cases (b) and (c) are similar for the given choice   of decay  parameters).  Thus, this channel may not play a significant role in revealing the Majorana phase dependence in the probability.

\begin{figure}[t!]
\includegraphics[width=3.5in]{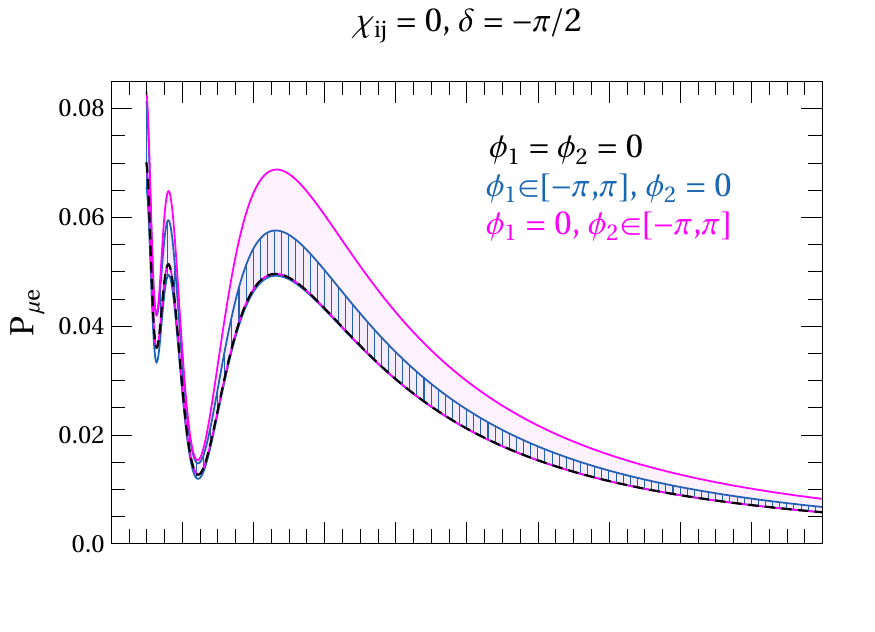}
\hskip -0.37in~~~
\includegraphics[width=3.1in]{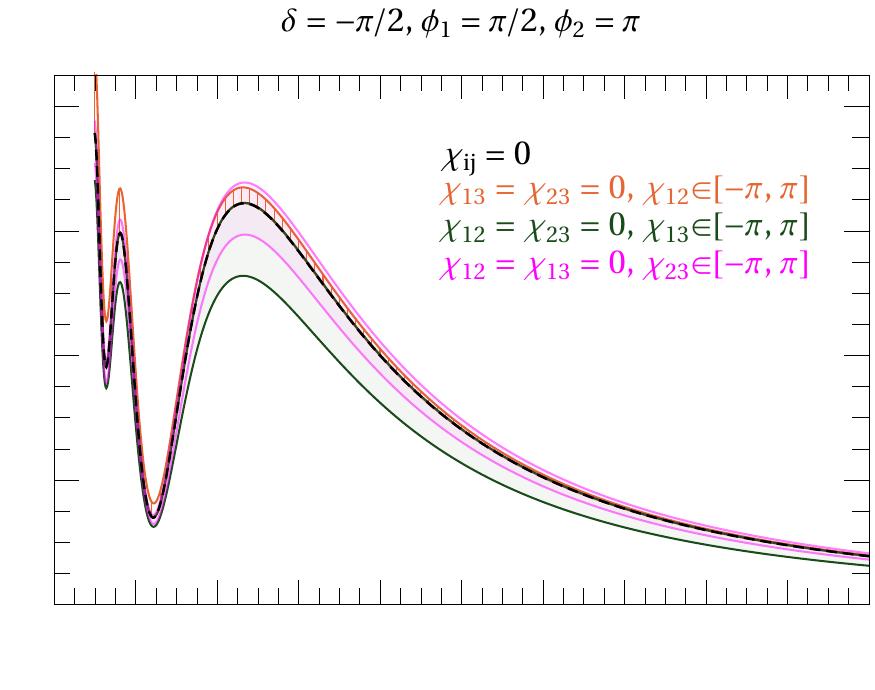}
\vskip -0.16in
\hskip -0.02in
\includegraphics[width=3.6in]{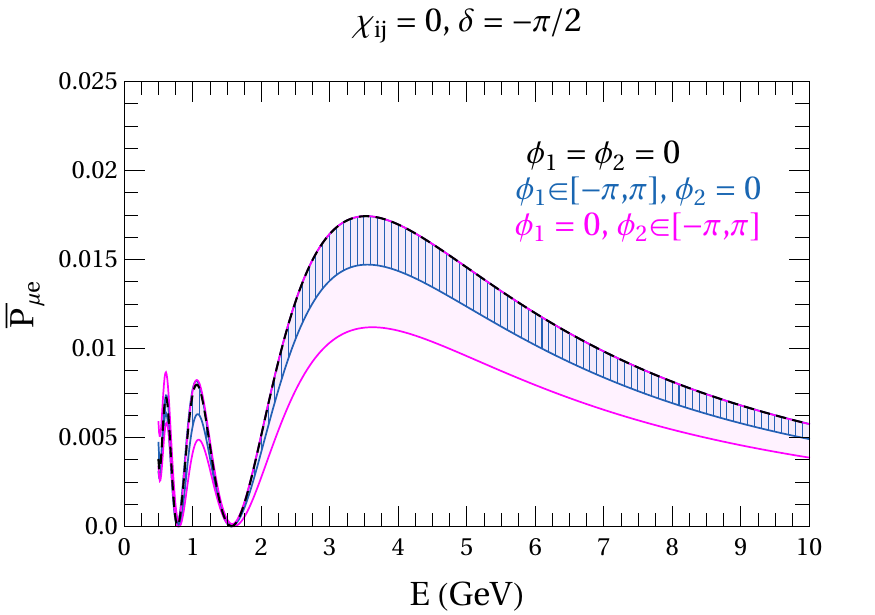}
\hskip -0.25in
\includegraphics[width=3.1in]{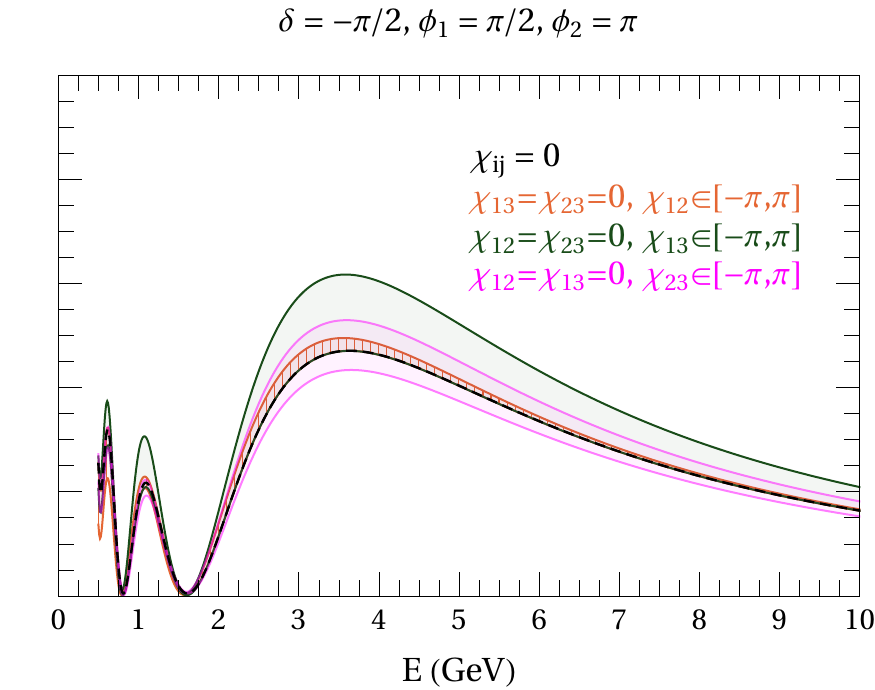}
\caption{\footnotesize Top (bottom) panel shows $P^{}_{\mu e}$ ($\bar{P}^{}_{\mu e}$) as a function of energy E. In the top left panel, black dotted curve corresponds to $\phi^{}_{1} = \phi^{}_{2} = 0$, whereas blue (magenta) band represents $\phi^{}_{1}\in[-\pi, \pi], \phi^{}_{2}=0$  ($\phi^{}_{1} = 0, \phi^{}_{2}\in[-\pi, \pi]$). In the top right panel, the black dotted curve is for  $\chi^{}_{ij} =0$, whereas the orange, green, and magenta bands correspond non-zero $\chi^{}_{12}, \chi^{}_{13}, $  and $\chi^{}_{23}$, respectively. Color coding is the same for the lower and top panels.
}
\label{fig:pmue_band}
\end{figure}

So far, we have considered fixed values of the  phases. We now vary the Majorana phases as well as the decay phases. In order to have clear understanding of the different phases, we vary one phase at a time and keep the other phases fixed. 
  Fig.~\ref{fig:pmue_band} shows that impact of varying Majorana phase as well as the decay phases on $P^{}_{\mu e}$ (top panel) and $\bar P^{}_{\mu e}$ (bottom panel)  for $L=1300$ km.  The plots on the left (right) depict the role of Majorana phases (decay phases)  for decay phases  set to zero (for  Majorana phases kept fixed). 
In order to examine the dependence of Majorana phases (top left panel of Fig.~\ref{fig:pmue_band}), we note that for $\chi^{}_{ij} =0\,(i,j = 1,2,3$ \textrm{and} $i \neq j)$ and $\delta=-\pi/2$, Eq.~\eqref{eq:prob_me} leads to
\bea
\label{eq:MajoranaPhase_mue_mm}
&&
P^{(\Gamma)}_{\mu e} (\phi^{}_{1} = \phi^{}_{2} = 0) \propto -  \gamma^{}_{13} c^{}_{12} - \gamma^{}_{23} s^{}_{12}\,,
 \\
&&
P^{(\Gamma)}_{\mu e} (\phi^{}_{1} \neq 0, \phi^{}_{2} =0) \propto -\gamma^{}_{13} c^{}_{12} - \gamma^{}_{23} s^{}_{12} \cos \phi^{}_{1} \,,
\\
&&
P^{(\Gamma)}_{\mu e} (\phi^{}_{1} =0, \phi^{}_{2} \neq 0) \propto - (\gamma^{}_{13} c^{}_{12} + \gamma^{}_{23} s^{}_{12} ) \cos \phi^{}_{2} \,.
\eea
\noindent
Thus, for $\phi^{}_{1} = \phi^{}_{2} = 0$, the probability has two terms with negative sign (this case is shown as black dotted line in Fig.~\ref{fig:pmue_band}, left panel); whereas, for $\phi^{}_{1} \neq 0, \phi^{}_{2} = 0$, with $\phi^{}_{1} \in [-\pi,\pi]$ leads to  second term changing sign, thereby enhancing the probability. The other case, $\phi^{}_{1} = 0, \phi^{}_{2} \neq 0$, with $\phi^{}_{2} \in [-\pi,\pi]$ leads to both terms changing sign, thereby enhancing the probability even more. We note that the band corresponding to  $\phi^{}_{2} \in [-\pi,\pi]$ (shown in magenta) is wider  in comparison to the band corresponding to  $\phi^{}_{1} \in [-\pi,\pi]$ (shown in blue).
At the location of first oscillation maximum for DUNE ($L=1300$ km), i.e., at $E \simeq 2.3$ GeV, $P^{(\Gamma)}_{\mu e} \sim 0.057$ for $\phi^{}_{1} = \phi^{}_{2} = 0$, 
$P^{(\Gamma)}_{\mu e} \in$ [0.057 - 0.066] for $\phi^{}_{1}  \in [-\pi,\pi], \phi^{}_{2} = 0$ and $P^{(\Gamma)}_{\mu e} \in$ [0.057 - 0.079] for $\phi^{}_{1} = 0, \phi^{}_{2}  \in [-\pi,\pi]$.

In order to examine the dependence of decay phases (top right panel of Fig.~\ref{fig:pmue_band}), we note that for $\delta^{}_{}=-\pi/2$ and  $\phi^{}_{1} = \pi/2$, $\phi^{}_{2} = \pi$, Eq.~\eqref{eq:prob_me} leads to
\bea
\label{eq:MajoranaPhase_mue_chi}
P^{(\Gamma)}_{\mu e} (\chi^{}_{ij} = 0) &\propto&  \gamma^{}_{13} c^{}_{12}  \,,
\\
P^{(\Gamma)}_{\mu e} (\chi^{}_{12} \neq 0) &\propto& 0 \,,
\\
P^{(\Gamma)}_{\mu e} (\chi^{}_{13} \neq 0) &\propto& \gamma^{}_{13} c^{}_{12} \cos \chi^{}_{13} \,,
\\
P^{(\Gamma)}_{\mu e} (\chi^{}_{23} \neq 0) &\propto& \gamma^{}_{13} c^{}_{12} 
- \gamma^{}_{23} s^{}_{12}  \sin \chi^{}_{23} \,.
\eea
So, for $\chi^{}_{i j} = 0$ $(i,j = 1,2,3$ \textrm{and} $i \neq j)$, the probability has only one term (this case is shown as black dotted line in Fig.~\ref{fig:pmue_band}, right panel).  Firstly, let us note that we have retained terms only up to $\mathcal O (\lambda^3)$  in Eq.~\eqref{eq:prob_me} and to this order, there is no contribution to $P^{(\Gamma)}_{\mu e}$  when $\chi^{}_{12} \in [-\pi,\pi]$. However, when $\chi^{}_{13} \in [-\pi,\pi]$,  due to the $\cos \chi^{}_{13}$ term, the probability is smaller in comparison to the case when $\chi_{ij} = 0$. For $\chi^{}_{23} \in [-\pi,\pi]$, we note that there are two terms and the second term has $\sin \chi^{}_{23}$ dependence so there is a spread around the curve corresponding to $\chi^{}_{ij} = 0$\,$(i,j = 1,2,3$ \textrm{and} $i \neq j)$.
At the location of first oscillation maximum for DUNE ($L=1300$ km), i.e., at $E \simeq 2.3$ GeV, 
$P^{(\Gamma)}_{\mu e} \sim 0.073$ for $\chi^{}_{ij} = 0$, 
$P^{(\Gamma)}_{\mu e} \in$ [0.06 - 0.073] for $\chi^{}_{13} \in [-\pi,\pi]$ and $P^{(\Gamma)}_{\mu e} \in$
[0.068 - 0.077] for $\chi^{}_{23}  \in [-\pi,\pi]$.
\begin{figure}[t!]
\includegraphics[width=3.3in]{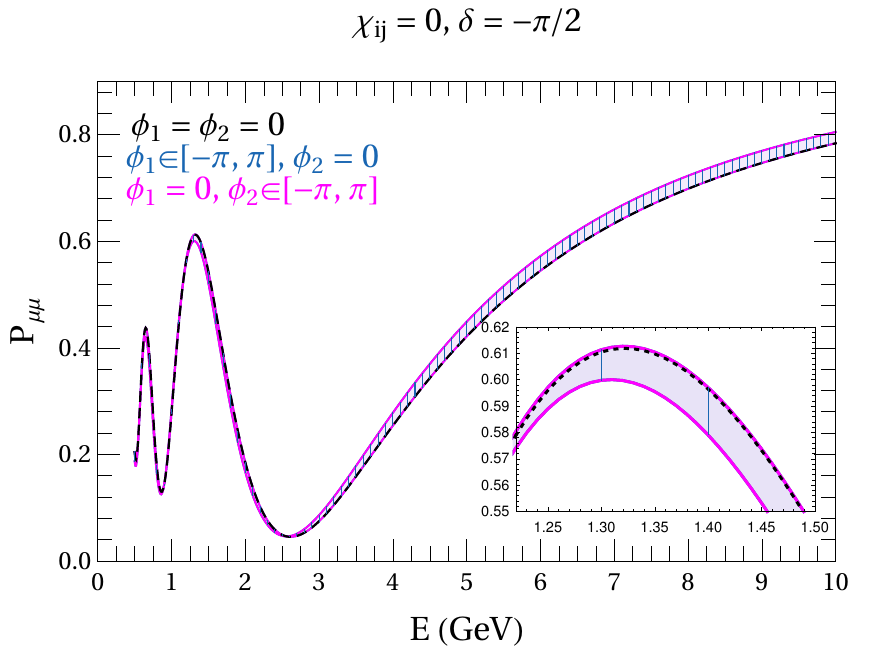}
\hskip -0.28in~~~
\includegraphics[width=3.3in]{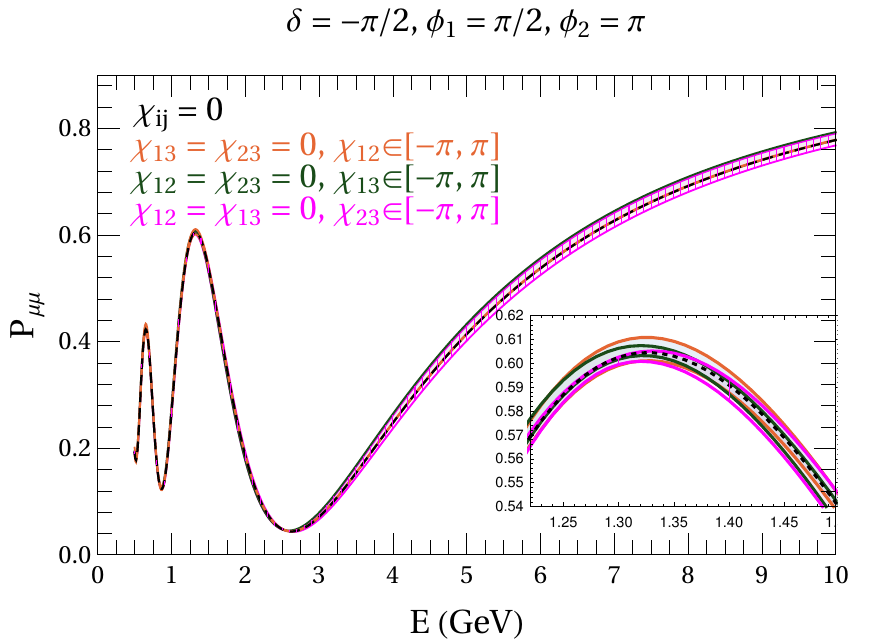}
\caption{\footnotesize Same as Fig.~\ref{fig:pmue_band} (top row) but for $\nu^{}_{\mu}\to\nu^{}_{\mu}$ channel.
}
\label{fig:pmumu_band}
\end{figure}
\begin{figure}[t!]
\hspace{-3cm} \includegraphics[width=9.5in]{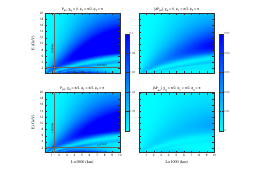}
\vspace{-1.3cm}
\caption{\footnotesize Oscillogram for $P^{}_{\mu e}$ (left panel) and $|\Delta P^{M}_{\mu e}| =|P^{\textrm{(w/ majorana)}}_{\mu e}- P^{\textrm{(w/o majorana)}}_{\mu e}|$ (right panel) in the energy $E$ versus baseline $L$ plane. The first column corresponds to probability in the presence of Majorana phases and the second column depicts the difference between the probabilities, with (w/) and without (w/o) Majorana phases for $\nu^{}_{\mu} \to \nu^{}_{e}$ channel. The value of Dirac CP phase is $\delta^{}_{} = - \pi/2$.} 
\label{fig:heatmap}
\end{figure}
The corresponding anti-neutrino probability, $\bar P^{}_{\mu e}$ for NH is plotted as a function of energy in the bottom panel of Fig.~\ref{fig:pmue_band}. In general, the phases and the matter term change sign (see Eq.~\eqref{eq:cpasym}) and one does not see large matter-induced enhancement over vacuum oscillations in the anti-neutrino channel ($\bar P^{}_{\mu e}$). We can understand the interplay of various phases in the case of $\bar P^{}_{\mu e}$ in similar manner as described above. Fig.~\ref{fig:pmumu_band} shows $P^{}_{\mu \mu}$ plotted as a function of energy for $L=1300$ km. For the muon disappearance channel, $P^{}_{\mu \mu}$, we note that the impact of Majorana phase variation is comparatively mild as can be seen also from Fig.~\ref{fig:prob_num_matt}.

So far we considered DUNE as a representative long baseline neutrino experiment and discussed the interplay of various phase terms appearing in the oscillation probability for $\nu_e$ appearance and $\nu^{}_\mu$ disappearance channels. In order to generalize this, we now consider a range of energies and baselines and depict the impact of phases in the form of oscillogram in two-dimensional plane of $E$ and $L$. We take $E$ = $0.5$ - $20$ GeV and $L$ = $100$ - $10000$ km. In order to quantify the effect of Majorana phases on the oscillation probability, we define a quantity 
\bea
 |\Delta P^M_{\alpha \beta}| &=& |P^{\textrm{(w/ Majorana)}}_{\alpha \beta}- P^{\textrm{(w/o Majorana)}}_{\alpha \beta}|\,,
 \label{eq:prob_diff2}
 \eea \noindent
 where $P^{\textrm{(w/ Majorana)}}_{\alpha \beta}$ is the probability with Majorana phases taken to be non-zero and $P^{\textrm{(w/o Majorana)}}_{\alpha \beta}$ is the probability with Majorana phases set to zero.

In Fig.~\ref{fig:heatmap}, we depict the oscillogram for $P_{\mu e}$ and $|\Delta P^M_{\mu e}|$. The top row exhibits the scenario when decay phases, $\chi^{}_{ij} =0$\,$(i,j = 1,2,3$ \textrm{and} $i \neq j)$ are set to zero. In the bottom row, we consider the case when $\chi^{}_{ij} =\pi/2\, (i,j = 1,2,3$ \textrm{and} $i \neq j)$. In the left panel, we plot the probability while in the right panel, we plot the $|\Delta P^M_{\alpha \beta}|$ given by Eq.~\eqref{eq:prob_diff2}. 
We note that the two Majorana phases have a larger impact on the probability in the vicinity of the first oscillation maximum of $P_{\mu e}$ at $E \simeq 2.3~\text{GeV}$ in the matter case, which is marked by a star in the left panel of Fig.~\ref{fig:heatmap}.
The impact of Majorana phases grow with $E$ and $L$.  
In the bottom panel of Fig.~\ref{fig:heatmap}, taking decay phases to be non-zero (for the considered choice of phases), we note that the phases  interfere destructively and hence the imprint of Majorana phases becomes weaker.
\subsection{CP asymmetry - impact of Majorana and decay phases}
\label{sec:CPasym}
\begin{figure}[t!]
\centering
\includegraphics[width=3.2in]{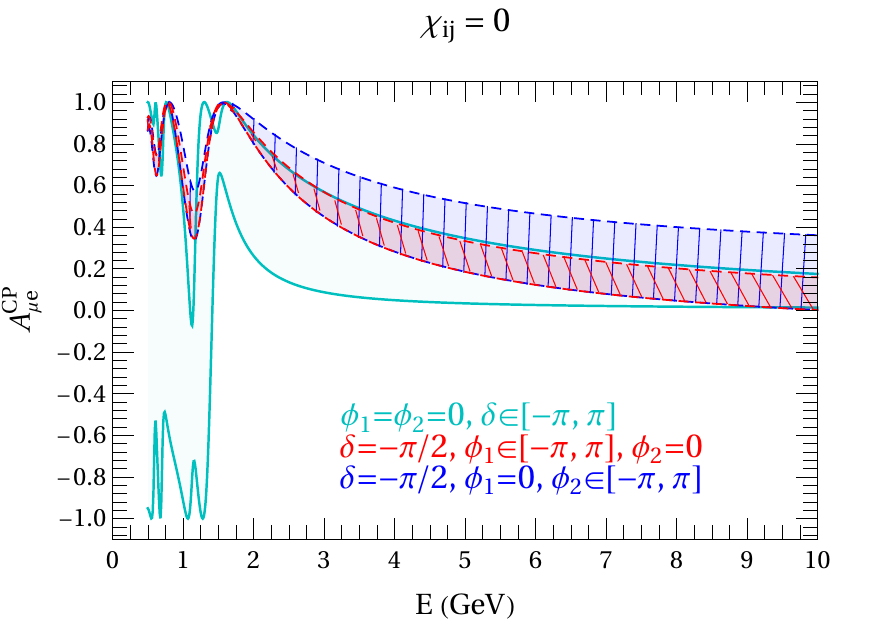}\hskip -0.28in~~~
\includegraphics[width=3.2in]{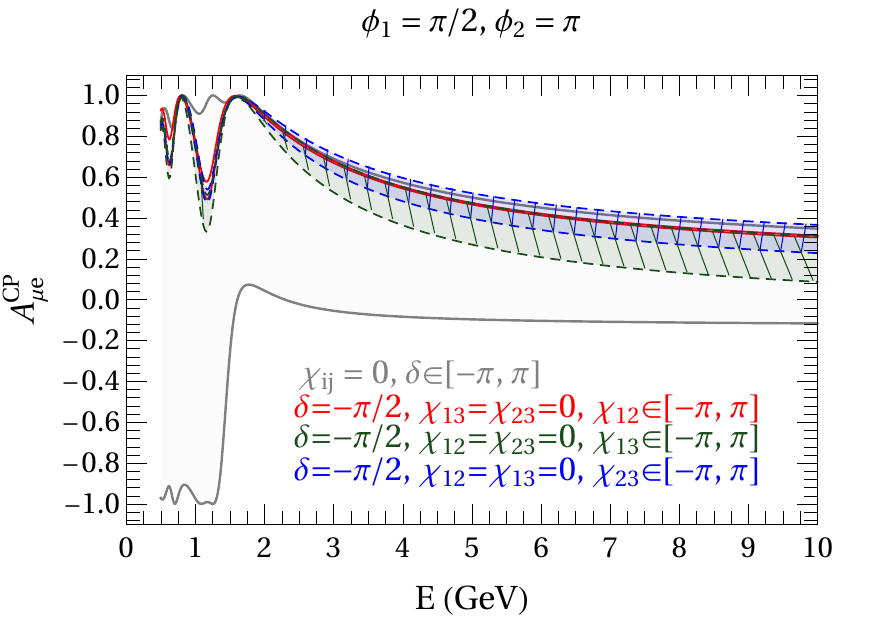}
\caption{\footnotesize $A^{CP}_{\mu e}$ as a function of energy E. Left panel: the cyan band corresponds to $\delta^{}_{}\in[-\pi, \pi]$ without decay, and the red (blue) band corresponds to non-zero $\phi^{}_{1}\,(\phi_{2})$ where  $\phi^{}_{1}\in[-\pi, \pi]$ ($\phi^{}_{2}\in[-\pi, \pi]$) with decay. Right panel: the red, green, and blue bands correspond to  $\chi^{}_{13} = \chi^{}_{23} = 0, \chi^{}_{12}\in[-\pi, \pi], \chi^{}_{12} = \chi^{}_{23} = 0, \chi^{}_{13}\in[-\pi, \pi]$, and $\chi^{}_{12} = \chi^{}_{13} = 0, \chi^{}_{23}\in[-\pi, \pi]$, respectively. The gray band represents the case with decay but for $\chi^{}_{ij} = 0$.
}
\label{fig:asym_majo}
\end{figure}

%
The CP asymmetry is defined as
{\small \bea
\label{eq:prob_diff3}
A^{CP}_{\alpha \beta} &=& \dfrac{\Delta P^{CP}_{\alpha \beta}}{P^{}_{\alpha \beta} + \bar{P}^{}_{\alpha \beta}}\, ,
\eea }\noindent
where $\Delta P^{CP}_{\alpha \beta} = P^{}_{\alpha \beta} - \bar{P}^{}_{\alpha \beta}$. In the general case of neutrino oscillations with decay considered in the present work,  $\Delta P^{CP}_{\alpha \beta} $ is given by (for the standard neutrino oscillations in matter, the CP asymmetry is given in~\cite{Masud:2015xva})
{\small\bea
 \label{eq:Acp}
\Delta P^{CP}_{\mu e}&=&4s^{2}_{13}s^{2}_{23}~\Theta^{}_{-}  - 8\alpha^{}_{}
  s^{}_{13} s^{}_{12} 
         s^{}_{23} c^{}_{12} 
         c^{}_{23}
\dfrac{\sin A \Delta}{A}  \Big[\Omega^{}_{-} \, \cos \Delta \cos \delta^{}_{} +  \Omega^{}_{+} \, \sin \Delta \sin \delta^{}_{}  \Big]\nonumber \\&-&8\gamma^{}_{3} \, \Theta^{}_{-}\, s^{2}_{13}s^{2}_{23}\Delta +4~\Theta^{}_{+}s_{13}s^{2}_{23}\Big\{\gamma^{}_{13} c^{}_{12} \sin[\delta^{}_{}-\phi^{}_{2}+\chi^{}_{13}]
+\gamma^{}_{23} s^{}_{12} \sin[\delta^{}_{}+\phi^{}_{1}-\phi^{}_{2}+\chi^{}_{23}]\Big\} \nonumber \\ &+& {\mathcal{O}} (\lambda^{4})\,,
\eea  }\noindent
where
{\small \bea
\Theta_{\pm}&=&\Bigg\{\dfrac{\sin^{2}(A-1)\Delta}{(A-1)^{2}} \pm \dfrac{\sin^{2}(A+1)\Delta}{(A+1)^{2}}\Bigg\}\,, \nonumber \\
\Omega_{\pm} &=& \Bigg\{\dfrac{\sin(A+1) \Delta}{A+1} \pm \dfrac{\sin(A-1) \Delta}{A-1}\Bigg\} \,.
\nonumber
\eea }
It may be noted that Eq.~\eqref{eq:Acp} depends on both extrinsic (matter-dependent) and intrinsic CP violating terms and there is a strong interplay of both kinds of contributions to the CP asymmetry.  As far as intrinsic contribution is concerned, there are three kinds of phases - Dirac phase ($\delta^{}_{}$), Majorana phases $\phi^{}_{i}$\,$(i=1,2)$ and decay phases $\chi^{}_{ij}$\,$(i,j = 1,2,3$ \textrm{and} $i \neq j)$ appearing in the above equation. 
In the absence of neutrino decay, Eq.~\eqref{eq:Acp} implies that the CP asymmetry depends only on the Dirac phase ($\delta^{}_{}$) and the matter-dependent terms~\cite{Masud:2015xva} and it is difficult to determine the source of CP violation in a clean manner~\cite{Rout:2017udo, Parveen:2023ixk}. 
In Fig.~\ref{fig:asym_majo}, we plot $A^{CP}_{\mu e}$ as a function of energy. In the left panel, we exhibit the dependence on Majorana phases for the case when decay phases  $\chi^{}_{ij} =0\,(i,j = 1,2,3$ \textrm{and} $i \neq j)$ are set to zero. Now, for the standard three flavor case, we expect to get non-zero CP asymmetry within the cyan band when $\delta$ is varied ($\delta \in [-\pi, \pi]$). For fixed $\delta^{}_{} = -\pi/2$, we can visualize the dependence on the Majorana phases $\phi_{i}$\,$(i=1,2)$ keep one non-zero at a time. The red (blue) band corresponds to $\phi^{}_{1} \in [-\pi, \pi]$ ($\phi^{}_{2} \in [-\pi, \pi]$) and $\phi^{}_{2} =0$ ($\phi^{}_{1} =0$). While the red band almost completely lies within the cyan band, the blue band is partially non-overlapping at $E \gtrsim 2$ GeV and can lead to larger CP asymmetry ($\simeq$ 0.35 - 0.82) in comparison to the standard three flavor case in matter.  In the right panel, we show the dependence on decay phases for certain choice of non-zero Majorana phases, $\phi^{}_{1}=\pi/2, \phi^{}_{2}=\pi$. We get non-zero CP asymmetry within the gray band for $\delta \in [-\pi, \pi]$ for vanishing decay phases $\chi^{}_{ij} =0\,(i,j = 1,2,3$ \textrm{and} $i \neq j)$ but non-vanishing $\gamma_{13}$ and $\gamma^{}_{23}$. For $\delta^{}_{} =-\pi/2$, the red, green and blue bands correspond to $\chi^{}_{12}  \in [-\pi, \pi]$, $\chi^{}_{13}  \in [-\pi, \pi]$ and $\chi^{}_{23}  \in [-\pi, \pi]$  respectively. We can note that there is not much dependence due to the variation of $\chi^{}_{12}$, since $\chi^{}_{12}$ does not appear in Eq.~\eqref{eq:Acp}  up to  ${\mathcal{O}}(\lambda^{3})$. 

\subsection{Impact of Majorana and decay phases on the parameter degeneracies}
\label{sec:deg}
The usual three flavor neutrino analysis of long baseline neutrino experiments is plagued by parameter degeneracies which can severely affect the determination of the standard oscillation parameters (such as the Dirac CP phase $\delta^{}_{}$, octant of $\theta^{}_{23}$, and the question of hierarchy i.e., the sign of $\Delta$)~\cite{Barger:2001yr, Minakata:2002qi}.  A given value of oscillation probability could result from more than one combination of parameters, thereby resulting in degeneracy and we need to devise appropriate strategies to distinguish the true solution from fake solutions~\cite{Barger:2001yr, Minakata:2002qi}. 
In the following we discuss how the parameter degeneracies are modified in presence of neutrino decay. As we have additional parameters $\gamma^{}_{ij}\,(i,j = 1,2,3$ \textrm{and} $i \neq j)$, $\chi^{}_{ij}\, (i,j = 1,2,3$ \textrm{and} $i \neq j)$ and $\phi^{}_{i}\,(i=1,2)$ impacting the oscillation probabilities, the discussion of degeneracy gets more involved. Here, we will focus on the  generalized $\textrm{sign}(\Delta)$ - $\theta_{23}$ - $\delta$ degeneracy impacting the appearance channel~\cite{Ghosh:2015ena} 
\bea
P^{}_{\mu e}[\Delta^{}_{},\theta^{}_{23}, \delta^{}_{}] &=&   P^{}_{\mu e}[-\Delta^{'}_{},\theta^{'}_{23}, \delta^{'}_{}]\,.  \label{eq:deg}
 \eea
One useful way to identify the degeneracies is to plot  $P_{\mu e}$ as a function of $\delta$ for a given baseline and energy as mentioned in~\cite{Ghosh:2015ena}. 
\begin{figure}[t!]
\includegraphics[width=3.55in]{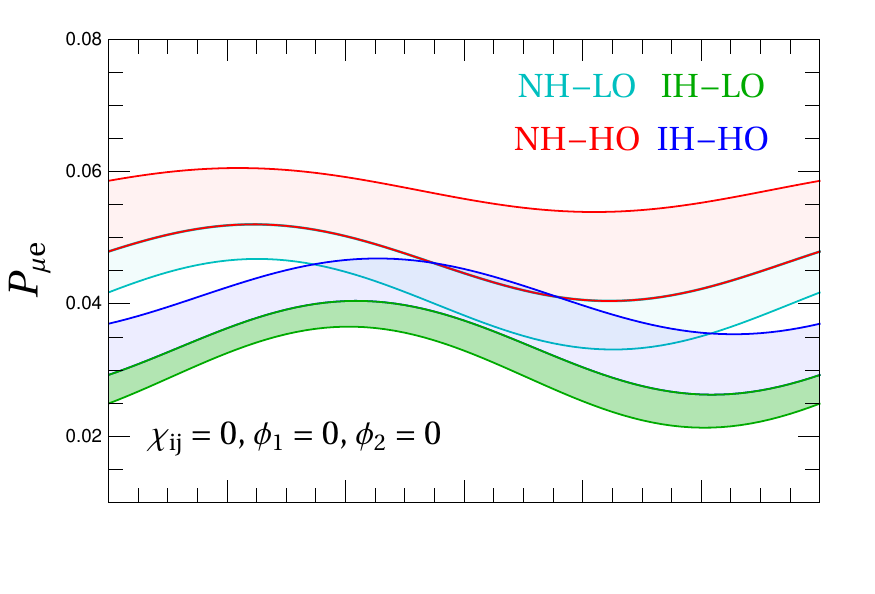}
\hskip -0.37in~~~
\includegraphics[width=3.2in]{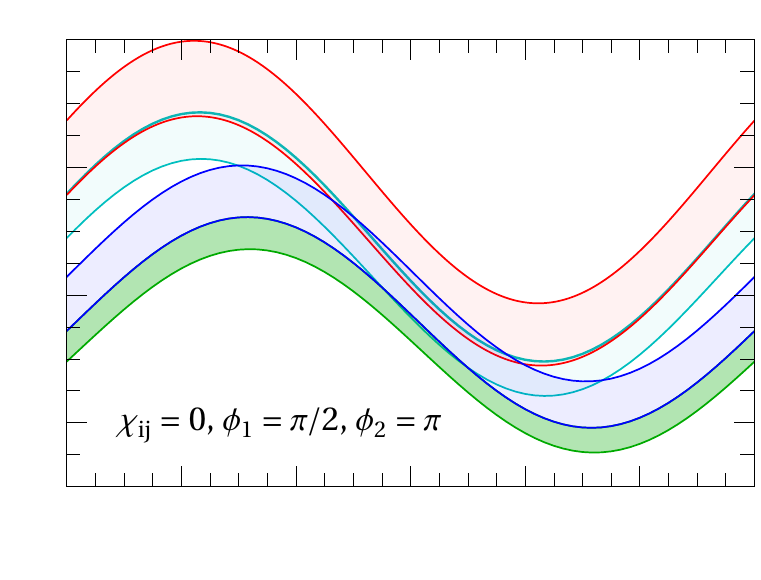}
\vskip -0.16in
\hskip -0.02in
\includegraphics[width=3.555in]{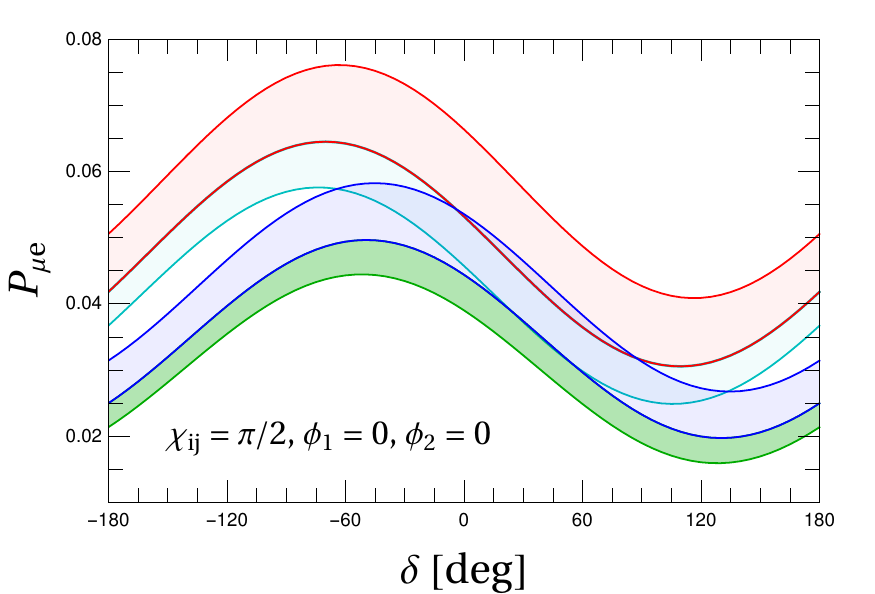}
\hskip -0.25in
\includegraphics[width=3.28in]{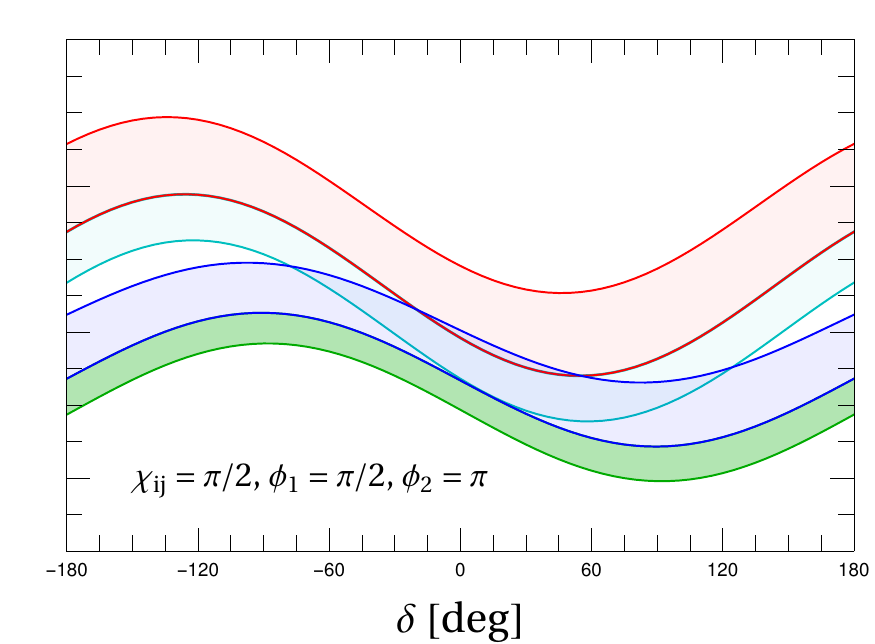}
\caption{\footnotesize First row: $P^{}_{\mu e}$ plotted as a function of $\delta^{}_{}$ for DUNE  ($E = 2.3$ GeV, $L=1300$ km) with zero and non-zero Majorana phases while setting the decay phases to zero.  Second row: Same as first row but with non-zero decay phases. The four bands correspond to different combinations of $\theta^{}_{23}$ octant (for LO, we take $\theta^{}_{23} \in$ [$41^{\circ}$ -  $44.5^{\circ}$] and for HO, we take $\theta^{}_{23} \in$ [$45.5^{\circ}$ - $52^{\circ}$]) and hierarchy (NH and IH). 
}
\label{fig:degeneracy}
\end{figure}
\begin{figure}[t!]
\includegraphics[width=6.5in]{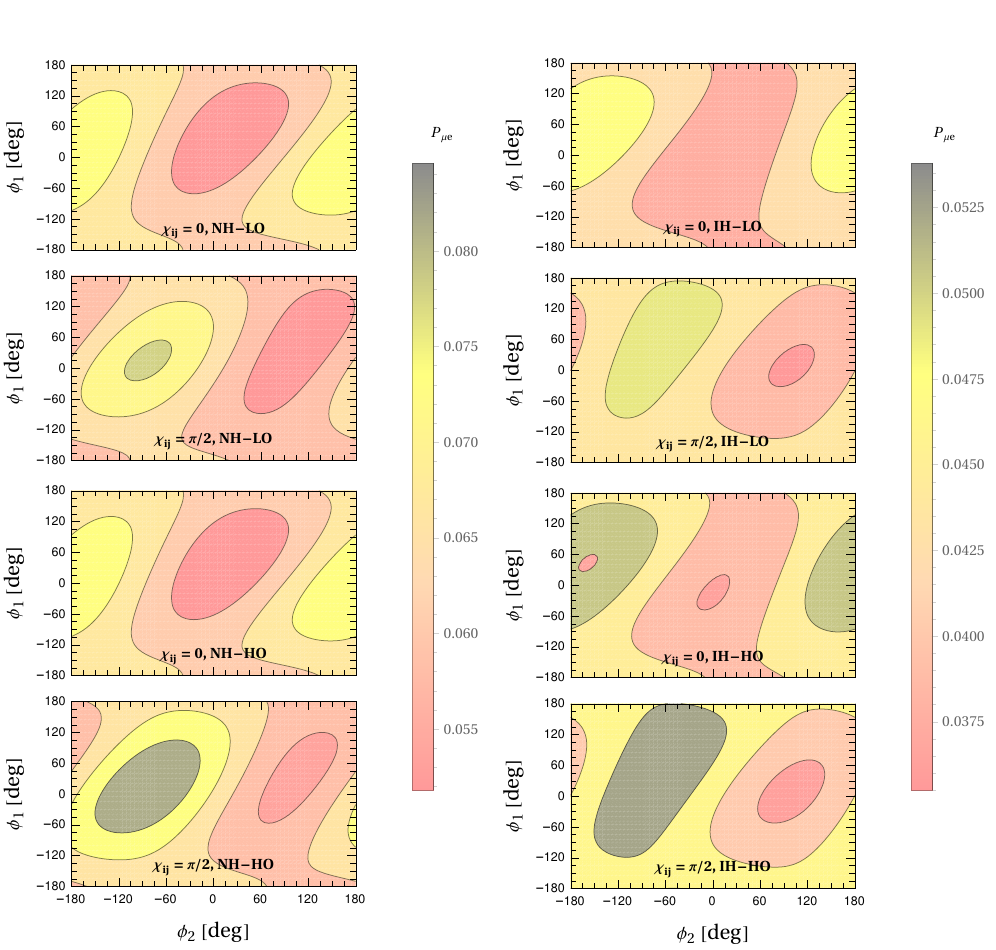}
\caption{\footnotesize Oscillogram for $P^{}_{\mu e}$, NH (left panel) and $P^{}_{\mu e}$, IH (right panel) in $\phi^{}_{1}$ - $\phi^{}_{2}$ plane for $\nu^{}_{\mu} \to \nu^{}_{e}$ channel for $\delta^{}_{} = - \pi/2$. The first and third row correspond to zero decay phases $(\chi^{}_{ij} = 0)$, whereas the second and fourth row correspond to non-zero decay phases  $(\chi^{}_{ij} = \pi/2)$. Also the top (bottom) two rows represent LO (HO). Here minimum (maximum) probability is shown using light-red (gray) color, where yellow color signifies intermediate probability.}
\label{fig:heatmap_VaryPhi}
\end{figure}
In Fig.~\ref{fig:degeneracy}, we plot the $P^{}_{\mu e}$ as function of $\delta^{}_{}$ in presence of neutrino decay for  DUNE  ($E=2.3$ GeV and $L=1300$ km).   The different bands correspond to combinations of hierarchy and octant\footnote{For lower octant (LO), we take $\theta^{}_{23} \in$ [$41^{\circ}$ - $44.5^{\circ}$], for higher octant (HO), we take $\theta^{}_{23} \in$ [$45.5^{\circ}$ - $52^{\circ}$].}: NH - LO, NH - HO, IH - LO, and IH - HO, represented by light cyan, red, green, and blue, respectively. 
The left panel corresponds to a representative case when we take the Majorana phases to be zero ($\phi^{}_{1}=\phi^{}_{2}=0$) while the right panel is for the case when Majorana phases are taken to be non-zero ($\phi^{}_{1}=\pi/2, \phi^{}_{2}=\pi$). From Fig.~\ref{fig:degeneracy}, we can identify the following degenerate solutions:
{\begin{enumerate}
\item[$\bullet$]
The cyan (NH - LO) and blue (IH - HO) bands overlap when $\delta \in$ [$-85 ^{\circ}$ - $130^{\circ}$]. This feature is present irrespective of the choice of Majorana phases $\phi^{}_{i}\,(i=1,2)$ (left and right panel) or for zero and non-zero values of $\chi^{}_{ij}\,(i,j = 1,2,3$ \textrm{and} $i \neq j)$ (top and bottom row).
We shall refer to this as the wrong hierarchy - wrong octant - right $\delta$ (WH - WO - R$\delta$)   degeneracy.
\item[$\bullet$]
The cyan band (NH - LO)  in the lower half plane (LHP) corresponding to $\delta \in [-180^{\circ} ,0^{\circ}]$ and the red band (NH - HO) in the upper half plane (UHP)  corresponding to $\delta \in [0^{\circ},180^{\circ}]$ give rise to same value of $P^{}_{\mu e}$. This corresponds to right hierarchy - wrong octant - wrong $\delta$ (RH - WO - W$\delta$) degenerate solution. This can be seen in the left and right panel of bottom row in Fig.~\ref{fig:degeneracy}. Similarly, the green (IH - LO) and blue (IH - HO) bands give rise to same value of $P^{}_{\mu e}$. This kind of degenerate solutions depend on the specific choice of decay parameters and Majorana phases. 
\item[$\bullet$]
Overall from Fig.~\ref{fig:degeneracy},  we could conclude that  the NH - HO scenario (red band) may avoid degeneracy challenges for the Dirac CP violating phase at approximately $-120^\circ$ (top right panel), and $-60^\circ$ (bottom left panel). A similar conclusion may be derived for the IH - LO scenario (green band) for $\delta^{}_{} \sim 120^\circ$.  On the other hand, NH - LO and IH - HO are likely to experience the most significant degeneracy issues.
\end{enumerate}}
In the discussion so far, we have taken fixed values of the Majorana phases, $\phi^{}_{i}\,(i=1,2)$ (see Fig.~\ref{fig:degeneracy}).  In order to visualize the effect of varying  $\phi^{}_{1}$ and $\phi^{}_{2}$, in Fig.~\ref{fig:heatmap_VaryPhi}, we present the iso-probability contours which depict degenerate regions in the plane of  $\phi^{}_{1}$ and $\phi^{}_{2}$ while keeping $\delta^{}_{}$ fixed at $-\pi/2$. Non-zero $\chi^{}_{ij}$ shifts the degenerate regions with respect to  $\chi^{}_{ij}=0$. Fig.~\ref{fig:heatmap_VaryPhi} allows us to understand the role played by the Majorana phases in the discussion of degeneracies. For instance, if we look at the bottom left panel for NH - HO case ($\chi^{}_{ij} = \pi/2$), we note that the probability is large in the region around $\phi^{}_{1}=0$ and $\phi^{}_{2} \sim -90^\circ$. However, if we take $\chi^{}_{ij} = 0$  (third row, left panel for NH - HO case), we note that this degeneracy is lifted. Additionally, there are degenerate solutions at specific values of $\phi^{}_{1}$ and $\phi^{}_{2}$. This implies that discussion of degeneracies in neutrino oscillation probabilities is further complicated due to the presence of Majorana phases and a careful assessment is needed.
\section{Conclusion}
\label{sec:conclusion}
Deciphering whether neutrinos are of Dirac-type or Majorana-type is of utmost importance in present day (astro-)particle physics and there have been several interesting proposals~\cite{Case:1957zza, Li:1981um, Kayser:1981nw, Kayser:1982br, Kim:2021dyj, Barger:2002vy, Pascoli:2002qm, deGouvea:2002gf, Pascoli:2005zb, Simkovic:2012hq, Adams:2022jwx, Dery:2024lem, Akhmedov:2024fsh}.  However, the smallness of neutrino mass poses a severe challenge. In the strict massless limit, Dirac and Majorana neutrinos are expected to be indistinguishable as they behave like Weyl fermions. 
Several current and upcoming neutrino oscillation experiments  \cite{NOvA:2016kwd, JUNO:2015zny, ESSnuSB:2013dql,T2K:2017hed, Hyper-KamiokandeProto-:2015xww, Hyper-Kamiokande:2016srs, Hyper-Kamiokande:2018ofw, MINOS:2020llm,Acciarri:2015uup, DUNE:2020ypp}
have the potential to explore physics beyond the SM such as neutrino decay  which could give rise to sub-dominant observational effects. The standard treatment of neutrino oscillations for any number of flavors ($\geq 2$) is insensitive to the Majorana phases appearing in the mixing matrix~\cite{Bilenky:1980cx, Schechter:1980gr, Doi:1980yb, Giunti:2010ec}. 
It turns out that a proper treatment of neutrino oscillations with decay naturally encompasses the possibility to have observable effects of Majorana phases at neutrino oscillation experiments, thereby opening up a new direction in order to understand the nature of neutrinos.  In the context of two neutrino flavors, it was shown that the oscillation probability has the imprint of the Majorana phase if the off-diagonal elements of the decay matrix were non-zero~\cite{Dixit:2022izn}. 
For the three flavor case with $3 \times 3$ mixing matrix containing both Dirac and Majorana phases, we present a complete treatment of neutrino oscillations with decay and establish conditions under which the Majorana phases  could leave their imprint at the level of detection probabilities of neutrinos. It may be pointed out that  the treatment presented in~\cite{Chattopadhyay:2021eba, Chattopadhyay:2022ftv}  considered the simplified form of mixing matrix with no (one) Dirac phase for the two (three) flavor case and did not take into account the Majorana phases.

We begin with the formalism  to describe three flavor neutrino oscillations along with decay (see Sec.~\ref{sec:formalism}). The effective Hamiltonian involves 9 new parameters ($\gamma^{}_{1}, \gamma^{}_{2}, \gamma^{}_{3}, \gamma^{}_{12}$, $\gamma^{}_{23}, \gamma^{}_{13}$, 
$\chi^{}_{12}, \chi^{}_{23}, \chi^{}_{13}$) in addition to the usual mass and mixing parameters - 3 angles ($\theta^{}_{12}, \theta^{}_{23}, \theta^{}_{13}$), Dirac phase ($\delta^{}_{}$), Majorana phases ($\phi^{}_{1},\phi^{}_2$) and 2 mass-squared splittings ($\Delta m^2_{21}$ and $\Delta m^2_{31}$). Incorporating Earth matter effects and using the Cayley-Hamilton formalism, we derive approximate analytical expressions for eigenvalues and oscillation probabilities  for neutrinos and anti-neutrinos. We highlight the dependence on these new parameters and in particular on the interplay of the three kinds of phases (Dirac, Majorana and decay)  giving rise to CP violating effects.  

Our results are contained in Sec.~\ref{sec:results}. We begin by assessing the accuracy of the analytical expressions with respect to numerical results in  Fig.~\ref{fig:prob_error}. For the electron appearance  ($\nu^{}_\mu \to \nu^{}_e$) channel, the agreement is fair and  the estimated measure of difference, $|\Delta P^{\textrm{error}}_{\mu e}| < 0.1\%$ for energies beyond $2$ GeV.  Similarly, for muon disappearance ($\nu^{}_\mu \to \nu^{}_\mu$) channel  and electron disappearance  ($\nu_e \to \nu_e$) channel, our results agree well ($|\Delta P^{\textrm{error}}_{\mu \mu} | < 0.1\%$ and $|\Delta P^{\textrm{error}}_{ee}| < 0.1\%$) for energies above $1.8$ GeV and $1.1$ GeV respectively.  We then depict the interplay of the Majorana phases and the decay terms  on the appearance and disappearance probability (see Figs.~\ref{fig:prob_num_matt}, \ref{fig:pmue_band} and \ref{fig:pmumu_band}) for a particular long baseline experiment, DUNE ($L=1300$ km). The probability difference given by Eq.~\eqref{eq:prob_diff2} is plotted as a oscillogram in the plane of energy and baseline in Fig.~\ref{fig:heatmap} and underscores the role of Majorana phase in some of the ongoing and upcoming experiments. 

While CP violation in the quark sector has been firmly established, CP violation in the leptonic sector still eludes us and this is one of the key questions being addressed by the ongoing as well as upcoming neutrino experiments. The presence of additional phases over and above the Dirac CP phase complicates the efforts in determining whether  CP symmetry is violated or not.  
The impact of the three phases on the CP asymmetry (in Eq.~\eqref{eq:Acp}) is shown in Fig.~\ref{fig:asym_majo}.
The presence of Majorana or decay phases leads to widening of asymmetry bands in general and a given value of asymmetry could arise from different combination of parameters as there are overlapping regions. In such a scenario, conclusively establishing CP violation at ongoing and future neutrino oscillation experiments would become even more challenging.
In addition to ascertaining whether CP is violated in the neutrino sector and determining the value of the CP phase $\delta$, there are two other unknowns, namely the question of neutrino mass hierarchy and the octant of $\theta^{}_{23}$. It is known that a certain combination of the parameters can lead to the same value of probability and this leads to parameter degeneracies. The structure of degeneracies is more complicated in presence of neutrino decay. In Fig.~\ref{fig:degeneracy}, we illustrate how certain values of the Majorana or decay phases could help resolve some parameter degeneracies for specific combinations of neutrino mass hierarchy and the octant of  $\theta^{}_{23}$ (see also Fig.~\ref{fig:heatmap_VaryPhi}). 

\medskip
\section*{Acknowledgments}
We thank Dibya S. Chattopadhyay for helpful discussions. SP acknowledges JNU for support in the form of fellowship. The numerical analysis has been performed using the HPC cluster at SPS, JNU funded by DST-FIST. NN is supported by the Spanish grants PID2023-147306NB-I00 and CEX2023-001292-S (MCIU/AEI/10.13039/501100011033), as well as CIPROM/2021/054 (Generalitat Valenciana). UKD  acknowledges support from the Anusandhan National Research Foundation (ANRF), Government of India under Grant Reference No. CRG/2023/003769. SP acknowledges IMSc, Chennai for support during the final stages of revision of this manuscript.
This research (SP and PM) was supported in part by the International Centre for Theoretical Sciences (ICTS) for participating in the program - Understanding the Universe Through Neutrinos (code: ICTS/Neus2024/04). The research of
PM has been partially supported by ANRF PAIR (ANRF/PAIR/2025/000029/PAIR-A).

\medskip

\noindent
{\bf{Note added :}} SP and SB  have contributed equally to this work.

\medskip
\appendix
\renewcommand{\theequation}{\thesection.\arabic{equation}}
\setcounter{equation}{0}
\renewcommand{\thesection}{\Alph{section}}

\section{Cayley-Hamilton formalism}
\label{sec:appA}
We give details of the Cayley-Hamilton formalism~\cite{
Ohlsson:1999xb, Ohlsson:2001vp} here. We wish to compute the time-evolution operator,
\bea
\label{S_matrix}
{\mathcal S} &=& \text{e}^{-i\mathcal{H} L}_{} \, ,
\eea \noindent
where $\mathcal{H}$ is the effective Hamiltonian in the flavor basis (Eq.~\eqref{eq:Gen_ham_dec}) and $L$ is  the propagation distance. 
Let $\mathcal{M} = -i{\mathcal{H}}L$ where $\mathcal{M}$ is an $N \times N$ matrix with eigenvalues $\lambda_a~(a=1,2,\ldots, N)$. The characteristic equation for ${\mathcal{M}}$ is given by,
\bea
 \det(\mathcal{M}-\lambda \boldsymbol{I}_N) = \lambda^{N} + a^{}_{N-1} \lambda^{N-1} + \ldots + a^{}_{1} \lambda + a^{}_0\boldsymbol{I}_N =  0\,,
\label{eq:C_eqn}
\eea \noindent
where, $\lambda$ is  an eigenvalue of $\mathcal{M}$, $\boldsymbol{I}$ is the $N \times N$ identity matrix and 
$a_n~(n=0,1,\ldots, N-1)$ are the coefficients referred to as the principal invariants. 
As every matrix satisfies its own characteristic equation, we can substitute $\lambda$ by $\mathcal{M}$ in Eq.~\eqref{eq:C_eqn} and obtain
\bea
\mathcal{M}^N  &=& - a^{}_{N-1} \mathcal{M}^{N-1} - \ldots - a^{}_{1} \mathcal{M} - a^{}_{0}\boldsymbol{I}_{N}\,.
\label{eq:M_eqn}
\eea
Moreover, for $p\geqslant N$,
 \bea
\mathcal{M}^{p}  &=&  a^{(p)}_{N-1} \mathcal{M}^{N-1}_{} + \ldots + a^{(p)}_{1} \mathcal{M} + a^{(p)}_{0}\boldsymbol{I}_{N}\,,
\label{eq:Mp_eqn}
\eea \noindent
where $a^{(p)}_{n} (n = 0, 1, \ldots N-1)$ are some coefficients. We note that the exponential of  matrix $\mathcal{M}$ given by
\bea
\text{e}^\mathcal{M} = \sum^{\infty}_{n=0} \frac{1}{n!} \mathcal{M}^{n}\,,
\eea
is an infinite series. Using Eq.~\eqref{eq:M_eqn} and Eq.~\eqref{eq:Mp_eqn}, we can express it as a finite series,
\bea
\text{e}^\mathcal{M} &=&  \sum^{N-1}_{n=0} c_n \mathcal{M}^n\,,
\label{eq:appA6}
\eea
with coefficients $c_{n}~(n=0,1 \ldots, {N-1})$ to be determined and $N$ being the dimension of ${\mathcal M}$.

Now for $N=3$ case, 
\bea
\text{e}^{-i\mathcal{H} L} &=& \phi e^{-i\mathcal{T}L} = \phi \left(c_{0}\boldsymbol{I}_{3} -i c^{}_{1} L\mathcal{T} - c_{2} L^2 \mathcal{T}^2 \right),
\label{eq:T}
\eea \noindent
where $\phi\equiv e^{-i (\text{tr} \mathcal{H}) L/3}$ is a complex phase factor, $\mathcal{T} \equiv {\mathcal H} -  \frac{1}{3} (\text{tr} \mathcal{H}) \boldsymbol{I}_3$ is a traceless matrix, i.e., $\text{tr} \mathcal{T}=0$.  and $\text{tr} \mathcal{H} = E^{}_{1} + E^{}_{2} + E^{}_{3} + V^{}_{\text{cc}}$. {Next, we can find the coefficients $c_0,c_1,c_2$ by expressing Eq.~\eqref{eq:T} in terms of the eigenvalues $\lambda_a (a=1,2,3)$ as follows
\bea
\text{e}^{-i\lambda_{a} L} &=& \left( c^{}_{0}\boldsymbol{I}_{3} -i c^{}_{1} L\lambda^{}_{a} - c^{}_{2} L^2_{} \lambda^2_{a} \right).
\nonumber 
\label{eq:lambda}
\eea
In matrix form, 
\bea
 \left(\begin{array}{l}\text{e}^{-i\lambda_{1} L}\\\text{e}^{-i\lambda_{2} L}\\ \text{e}^{-i\lambda_{3} L} \end{array}\right)
&=&
\left(\begin{array}{lll} 1 &\, -i  L\lambda_{1} &\,- L^2 \lambda^2_{1} \\1&\,-i L\lambda_{2}&\, - L^2 \lambda^2_{2} \\ 1&\, -i L\lambda_{3}&\, - L^2 \lambda^2_{3} \end{array}\right) \left(\begin{array}{l}c_0\\c_1 \\c_2 \end{array}\right) \,.
\eea
Finally, we can then insert the form of $c^{}_{0},c^{}_{1},c^{}_{2}$ in \eqref{eq:T}, to get 
\bea
\text{e}^{-i\mathcal{H}L}&=&\dfrac{1}{(\lambda_1-\lambda_2)(\lambda_1-\lambda_3)} \, \phi \, \text{e}^{-i\lambda_1 L}\,\left[\mathcal{T}^{2}-(\lambda_2+\lambda_3)\mathcal{T}+\lambda_2\lambda_3\boldsymbol{I}\right]\,\nonumber \\
&+&\dfrac{1}{(\lambda_2-\lambda_1)(\lambda_2-\lambda_3)}\, \phi \, \text{e}^{-i\lambda_2 L}\big[\mathcal{T}^{2}-(\lambda_1+\lambda_3)\mathcal{T}+\lambda_1\lambda_3\boldsymbol{I}\big]\nonumber \\
&+&\dfrac{1}{(\lambda_3-\lambda_1)(\lambda_3-\lambda_2)} \, \phi \, \text{e}^{-i\lambda_3 L}\big[\mathcal{T}^{2}-(\lambda_1+\lambda_2)\mathcal{T}+\lambda_1\lambda_2\boldsymbol{I}\big].
\eea }

\section{Event rate analysis and bounds on neutrino decay} 
\label{sec:appB}
\noindent
In order to generate the event rates, we perform simulations using the software General Long Baseline Experiment Simulator (GLoBES)~\cite{Huber:2004ka,Huber:2007ji}.
For these simulations, we use the configuration files from the DUNE Technical Design Report (TDR)~\cite{DUNE:2020ypp,DUNE:2021cuw}.
DUNE consists of an on-axis $40$~kt liquid argon FD located at the Homestake Mine in South Dakota, with a baseline of $1300$~km.
A near detector (ND) with a target mass of $0.067$~kt will be installed at a baseline of $0.570$~km at the Fermi National Accelerator Laboratory (FNAL) in Batavia, Illinois. We utilize the standard (low energy) beam tune used in the DUNE TDR~\cite{DUNE:2020ypp}.
\begin{figure}[ht!]
\hskip -0.3in
\centering
\includegraphics[width=7in]{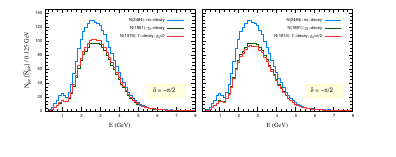}
\vspace{-1.5cm}
\caption{\footnotesize{The $\nu_{\mu}$ $\to$ $\nu_{e}$ ($\bar{\nu}_{\mu}$ $\to$ $\bar{\nu}_{e}$) event spectra at DUNE with no-decay (blue curve), $\gamma^{}_{3}$-decay (green curve), $\Gamma$-decay with $\chi^{}_{ij}=0$ (red curve, left panel) and $\Gamma$-decay phases with $\chi^{}_{ij}=\pi/2$ (red curve, right panel).
A total runtime of $13$ years with $6.5$ years in $\nu$-mode and $6.5$ years in $\bar{\nu}$-mode has been considered.}} 
\label{fig:event}
\end{figure}
$N_{\alpha \beta}$ is given by
\bea
N_{\alpha \beta}(L) &=& N_{\text{target}} \int
\Phi_{\nu_\alpha} (E,L)\, P_{\alpha \beta} (E,L)\, \sigma_{\nu_\beta} (E)\, dE\,,
\label{eventeqn}
\eea \noindent
where $N_{\text{target}}$ denotes the number of target nucleons per kiloton of detector fiducial volume, and $P_{\alpha \beta}(E,L)$ is the oscillation probability in matter.
For the DUNE, $N^{\text{DUNE}}_{\text{target}} = 6.022 \times 10^{32}~N/\textrm{kt}$, where $N$ represents the number of nucleons.
Here, $\Phi_{\nu_\alpha}(E,L)$ is the $\nu_\alpha$ flux, and $\sigma_{\nu_\beta}(E)$ is the CC cross section of $\nu_\beta$.
For antineutrinos, the substitutions $N_{\alpha \beta} \to \bar N_{\alpha \beta}$, $\nu_\alpha \to \bar \nu_\alpha$, and $\nu_\beta \to \bar \nu_\beta$ are made~\cite{Bass:2013vcg, Masud:2016nuj}.

 We have generated the event spectra for the $\nu_\mu \to \nu_e\,(\bar{\nu}_{\mu} \to \bar{\nu}_{e})$ channels at DUNE for the following cases: (i) no-decay (blue curve), (ii) no-decay with only $\gamma^{}_{3}$-decay (green curve), (iii) no decay with $\gamma_3$-decay and $\Gamma$-decay with $\chi^{}_{ij}=0$ (red curve in the left panel), and (iv) no-decay with $\gamma^{}_{3}$-decay and $\Gamma$-decay with $\chi^{}_{ij}=\pi/2$ (red curve in the right panel). These are shown in Fig.~\ref{fig:event}. 
We  observe a reduction in the event rates when decay is considered. This is consistent with the conclusion drawn at the level of probability (see Sec.~\ref{sec:Majorana}).

Next, we carry out $\chi^{2}$ analysis to place constraints on invisible neutrino decay. The statistical $\chi^{2}$ function is defined as
\bea
\chi^2 &=& 2 \sum_{y}^{\text{flux}}\sum_{x}^{\text{mode}}\sum_{j}^{\text{channel}}\sum_{i}^{\text{bin}}\left[N_{ijxy}^{\rm test} - N_{ijxy}^{\rm true} - N_{ijxy}^{\rm true} \ln \left( \frac{N_{ijxy}^{ \rm test}}{N_{ijxy}^{\rm true}} \right) \right],
\label{chi}
\eea
where $N^{\rm true}$ and $N^{\rm test}$ denote the predicted event rates for the true and test hypotheses, respectively. The index $i$ runs over the reconstructed energy bins in the range $0$–$20$ GeV~\footnote{In the present analysis, we use a total of $62$ energy bins in the range $0.5$–$10$ GeV: $62$ bins of width $0.125$ GeV in the interval $0.5$–$8$ GeV and $2$ bins of width $1$ GeV in the interval $8$–$10$ GeV~\cite{DUNE:2021cuw}.}.
The index $j$ labels the oscillation channels ($\nu_{\mu}\to\nu_{e}$ and $\nu_{\mu}\to\nu_{\mu}$), while $x$ runs over the neutrino and antineutrino running modes ($\nu$ and $\bar{\nu}$) and $y$ accounts for the different flux configurations used. The true values of the oscillation parameters and their uncertainties are given in Table~\ref{tab:parameters}, and all relevant oscillation parameters are marginalized in the analysis. Systematic uncertainties are incorporated using the pull method~\cite{Gonzalez-Garcia:2004pka, Fogli:2002pt}, with overall normalization errors included separately for signal and background~\cite{DUNE:2021cuw}.

\begin{figure}[t!]
\hskip -0.3in
\centering
\includegraphics[width=4in]{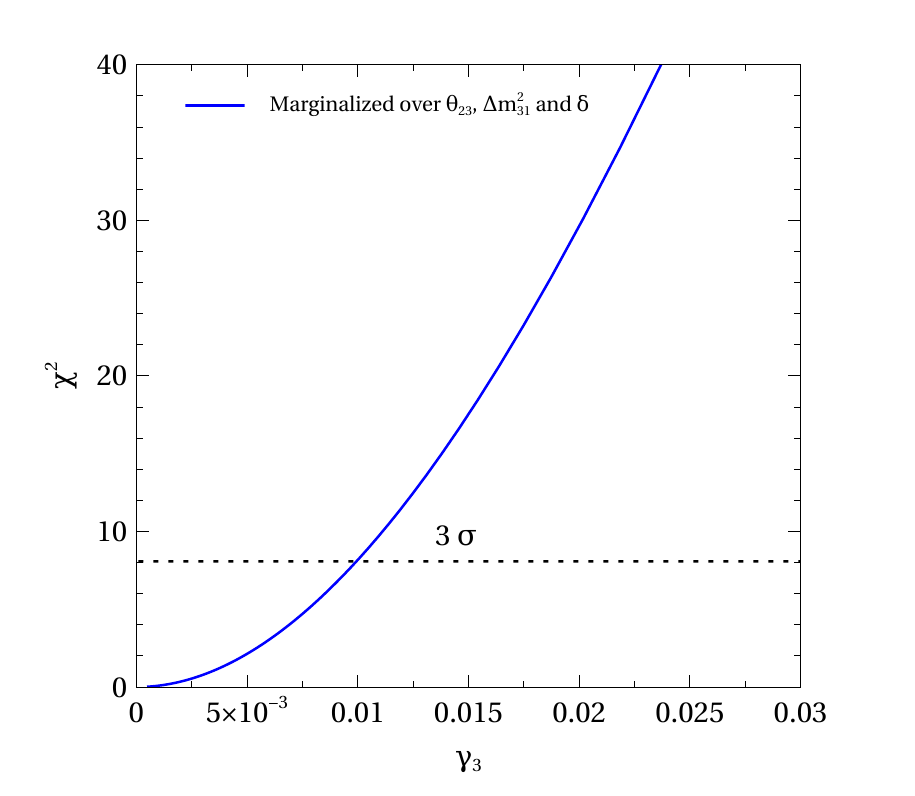}
\caption{\footnotesize{$\chi^{2}$ as a function of $\gamma_3$ for DUNE, marginalized over $\theta^{}_{23}$, $\ldm$ and $\delta$. The data has been generated assuming no-decay scenario.}}
\label{fig:bound}
\end{figure}

In Fig.~\ref{fig:bound}, we present the $\chi^{2}$ sensitivity as a function of the decay parameter $\gamma_{3}$, with marginalization over $\theta{}_{23}, \ldm$ and $\delta$. To obtain these results, we consider both appearance and disappearance channels in $\nu$ and $\bar{\nu}$-modes with variation $\delta\in[-\pi, \pi]$. The bound on $\gamma_3$ using the configuration of DUNE is consistent with~\cite{Dey:2024nzm}.
\bibliographystyle{unsrt}
\bibliography{ref}
\end{document}